\documentclass[
reprint,
 amsmath,amssymb,
 aps,
]{revtex4-2}

\usepackage{graphicx}% Include figure files
\usepackage{dcolumn}% Align table columns on decimal point
\usepackage{bm}% bold math
\usepackage{multirow} % for \multirow in tables
\usepackage{subcaption}
\usepackage[hidelinks]{hyperref}% add hypertext capabilities
\usepackage{comment}
\usepackage{ragged2e} 

\usepackage{caption}
\usepackage{booktabs}
\usepackage{float}

\usepackage[parfill]{parskip}
\begin{document}

\preprint{APS/123-QED}

\title{Jet Quenching Identification via Supervised Learning in Simulated Heavy-Ion Collisions}% Force line breaks with \\

\author{Leonardo Lima da Silva}
 \email{llsilva@if.usp.br}
\author{Marcelo Gameiro Munhoz}%
 \email{munhoz@if.usp.br}
\affiliation{%
 Instituto de Física, Universidade de São Paulo\\
}%


%  \homepage{http://www.Second.institution.edu/~Charlie.Author}
% \affiliation{
%  Second institution and/or address\\
%  This line break forced% with \\
% }%
% \affiliation{
%  Third institution, the second for Charlie Author
% \affiliation{%
%  Authors' institution and/or address\\
%  This line break forced with \textbackslash\textbackslash
% }%

% \collaboration{CLEO Collaboration}%\noaffiliation

\date{\today}% It is always \today, today,

\begin{abstract}
Jet modification in heavy-ion collisions provides microscopic access to the properties of the quark-gluon plasma. However, conventional approaches based on traditional global observables, such as \(R_{AA}\), capture limited information about the complex dynamics of parton-medium interactions during hard scatterings. In this work, we apply sequential machine learning architectures to the jet declustering history tree, achieving improved classification performance compared with static models that learn only from a single stage of the jet evolution. We find that models trained on different medium implementations exhibit meaningful performance modification under cross-domain validation, indicating that machine learning is sensitive to implementation-specific features that traditional observables may not resolve. These results suggest new opportunities for using machine learning as an analysis tool to overcome some of the limitations of traditional jet-modification studies.

%\begin{description}
%\end{description}
\end{abstract}

\maketitle

%\tableofcontents

\section{\label{sec:level1} Introduction}
A high-energy collision of two nuclei, whether in a fixed-target or a collider setup, produces numerous new particles. This output exceeds what one might expect from individual nucleon-nucleon collisions. It has been hypothesized that at high temperatures and densities, hadronic matter transitions to a state of free quarks and gluons, known as the \textit{quark-gluon plasma} (QGP) \cite{heinz2000evidence, adams2005experimental}. %This led to the development of Quantum Chromodynamics (QCD) \cite{gross202350} and more accurate predictions through lattice calculations (LQCD) \cite{gupta1998introduction}.

Jets provide direct experimental access to QGP properties through their modification during medium propagation \cite{cao2021jet}. Traditionally, these effects are observed using a collection of jets from heavy-ion collisions. One example is the nuclear modification factor, $R_{AA}$, that measures the suppression of particle production in the presence of a dense medium \cite{monalisa2024}.

Early studies of inclusive suppression observables showed that the extracted energy-loss coefficient depends on the hydrodynamic medium used to describe the QGP evolution~\cite{PhysRevC.75.031902}. In the present work, we revisit this medium-dependence from a jet-by-jet perspective using supervised machine-learning classifiers.

Jets are typically defined as a succession of partonic emissions initiated by the scattering of these particles at high energies. Due to confinement, these partons undergo a fragmentation process, forming a collimated set of partons, which subsequently transform into hadrons through a hadronization process and can be experimentally identified. 

The interaction of partons with the medium occurs through energy loss processes, such as elastic collisions or gluon emission, and is known as \emph{jet quenching} \cite{cao2021jet}. As mentioned, the comparison between a collection of jets produced in heavy-ion collisions and those produced in proton-proton collisions is a common way of studying the QCD medium. However, one can wonder about the impact of studying the influence of the QGP using a jet-by-jet approach.

Machine learning is particularly well-suited to learning statistical patterns from individual jets and making predictions based on their corresponding labels. Thus, it is a natural candidate as a tool for studying jet quenching on a jet-by-jet basis.
%High-energy quarks and gluons passing through the quark-gluon plasma experience differential energy loss due to elastic scattering with the plasma constituents. This energy loss is experimentally detected through the jet suppression of energy and transverse momentum distributions in comparison to those observed in proton-proton (pp) reference collisions.

In recent years, the field of high-energy physics (HEP) has experienced a significant resurgence of interest in the application of Machine Learning \cite{singh2007review}. Previous studies have explored deep-learning strategies for jet-quenching classification, showing that neural-network architectures can learn enough information from jet-level representation to discriminate between medium and vacuum-like jets \cite{Apolinario:2021olp}. In addition, comprehensive reviews have highlighted the growing applications of ML techniques specifically for jet quenching studies \cite{du2023overview}. 

Other approaches have explored energy-flow-network-based representations for jet-quenching classification, combining constituent-level learning with physics-motivated observables \cite{Goncalves:2025asw}.

Recent experimental work at ALICE has demonstrated the practical application of ML techniques for differential jet quenching measurements \cite{bossi2023novel}. It is essential to distinguish between jets generated within this medium and those originating from vacuum conditions to interpret the dynamics involved in these collisions accurately. 

In this manuscript, we apply various supervised machine learning algorithms in the task of jet quenching classification using two hydrodynamical models to study the effect of the medium on quenching effects. We begin with a simple medium description by \textsc{Jewel} \cite{zapp2013perturbative, zapp2014jewel}, which we refer to as \emph{Default}. Then, we use \textsc{Jewel} coupled with a more realistic environment named \emph{viscous Ultrarelativistic Smoothed Particle hydrodynamics } or, simply, v-USPhydro \cite{Noronha-Hostler:2013hsa}. We elaborate more about these two medium frameworks in the following sections. For jet classification, we use five different ML models: Random Forest \cite{breiman2001random}, Multilayer Perceptrons (MLP) \cite{rumelhart1986learning}, Long Short-Term Memory (LSTM) \cite{hochreiter1997long}, LSTM coupled with an attention mechanism \cite{bahdanau2014neural}, and Transformers \cite{vaswani2017attention}.

This work is systematically organized as follows: Section II establishes the simulation framework within \textsc{Jewel} and characterizes the jet substructure observables utilized as machine learning input features. Section III provides a brief description of the neural network architectures employed in our classification analysis. Section IV presents our results, distinguishing between performance achieved using static (non-sequential) jet substructure features versus sequential jet grooming history as input variables.

Our analysis shows that models trained on sequential jet data significantly outperform those trained on static data. This suggests that understanding jet modification as a temporal process is more effective. Additionally, the models exhibit considerable sensitivity to the details of medium implementation, indicating that machine learning can differentiate between medium frameworks that yield similar results in traditional observables, such as the $R_{AA}$.

\section{Simulation Setup and Data Preparation}

Jet Evolution With Energy Loss, commonly known as \textsc{Jewel} \cite{zapp2013perturbative, zapp2014jewel}, is a Monte Carlo framework for simulating QCD jet evolution in heavy-ion collisions. It models the interplay between QCD radiation and parton-medium interactions using a fully microscopic perturbative approach with minimal assumptions. 

\textsc{Jewel} incorporates \textsc{Pythia} \cite{sjostrand2006pythia, sjostrand2015introduction} as the underlying framework for event generation, leveraging its models and formalisms to simulate processes such as parton showers, hadronization, and decays. This allows \textsc{Jewel} to complement its detailed description of medium-induced effects on jets, including phenomena such as energy loss and radiative coherence.

\textsc{Jewel} models the medium created in heavy-ion collisions in a simplified manner. The medium consists of a static cloud of thermal partons with momentum distributions following a thermal profile and density evolving according to the Bjorken expansion \cite{zapp2014jewel}, which assumes longitudinal expansion only. Jet-medium interactions are implemented via elastic scattering and medium-induced radiation, allowing the simulation of parton energy loss while traversing the quark-gluon plasma (QGP). The initial spatial distribution of the medium is determined using Glauber’s model \cite{miller2007glauber}. Although \textsc{Jewel} does not include a realistic hydrodynamic evolution of the QGP, it allows control over key parameters such as the initial temperature, formation time, and the presence of recoiling partons. The energy loss is computed through perturbative processes, combining elastic scatterings and medium-induced gluon radiation. The latter follows the Baier–Dokshitzer–Mueller–Peigné–Schiff–Zakharov (BDMPS-Z) formalism, which incorporates the Landau–Pomeranchuk–Migdal (LPM) effect, and accounts for multiple scatterings within the medium \cite{baier1997radiative, zakharov1997radiative}. For more details, one can refer to \cite{kunnawalkam2017medium}.

To overcome the limitations of this simplified medium, \textsc{Jewel} can be coupled to more realistic hydrodynamic models. In this work, we employ the v-USPhydro description of hydrodynamics, which solves the relativistic hydrodynamic equations using the Smoothed Particle Hydrodynamics (SPH) method \cite{liu2010smoothed}. This model simulates the time evolution of the QGP in $2+1$ dimensions, including viscous effects and pressure gradients, with initial conditions provided by the Reduced Thickness Event-by-event Nuclear Topology ($\mathrm{T_R ENTo}$) model \cite{Moreland:2014oya}. The hydrodynamic evolution uses an equation of state based on lattice-QCD results at high temperatures, smoothly matched to a hadron-resonance-gas description at lower temperatures. In the present setup, we use the WB21/PDG16+ parametrization \cite{Alba_2017, Bors_nyi_2021}, following the v-USPhydro configuration adopted in previous phenomenological studies \cite{bernhard2019bayesian}. 

By embedding \textsc{JEWEL} jets into this dynamically evolving medium, one obtains a medium description that includes transverse expansion, local flow, viscous hydrodynamic evolution, and event-by-event fluctuations of the QGP. 
The \textsc{JEWEL}+v-USPhydro implementation used in this work follows from the work in Ref.~\cite{Barreto:2022ulg}.

We employ these two complementary medium implementations to probe sensitivity to QGP evolution details. In summary, the simplified thermal description of \textsc{Jewel} enables controlled studies of basic jet-medium interactions, whereas v-USPhydro's realistic hydrodynamic evolution includes anisotropic expansion and viscous effects characteristic of actual heavy-ion collisions.

The results presented in this manuscript were obtained using \textsc{Jewel} 2.4.0, which incorporates a slightly modified version of \textsc{Pythia} 6.4.25. The parameters chosen for the simulations follow those used in Ref.~\cite{crispim2024jet}, except that in this work, we adopt a broader transverse momentum range of $p_T \in [40,\, 450]$~GeV to investigate the impact of $p_T$ on the quenching effect. All events simulated for this work were performed with recoils on, which means that the medium particles that undergo interactions with the jets are included in the jet reconstruction. The treatment of recoils and background subtraction in JEWEL is known to affect jet substructure observables and has been discussed in detail in \cite{Milhano:2022kzx}. We do not explore the effects of recoils on observables in this manuscript. For more information, refer to \cite{monalisa2024}.

The observables analyzed in this study were computed using the code provided in \cite{crispim2024jet}, with a few modifications to suit our specific goals, and we refer to the implementation in the section Data Availability. We categorize the observables into two main groups, \textit{static} and \textit{sequential}. The static, or non-sequential, results are computed from the first splitting that satisfies the Soft Drop condition \cite{larkoski2014soft}, representing a snapshot of the declustered jet. In contrast, \textit{sequential} observables are extracted throughout the entire declustering process, associating a value with each step of the jet evolution and thus capturing its dynamical development. We elaborate more about these observables in Section \ref{sec:observables}.

Finally, we divide the resulting dataset into several intervals of transverse momentum to study the quenching effect across different regions of the $p_T$ spectrum. In this work, we study three intervals of transverse momentum,  namely $[40, 60]$, $[80, 250]$, and $[200, 400]$ GeV.

%Although we employ \textsc{Jewel} in this investigation, alternative Monte Carlo event generators for relativistic heavy-ion collisions include HYDJET++ \cite{lokhtin2009heavy}, Q-PYTHIA \cite{armesto2009q}, MARTINI \cite{schenke2009martini}, HIJING \cite{gyulassy1994hijing}, and THERMINATOR \cite{kisiel2006therminator}. Several of these frameworks incorporate thermal background effects alongside hard and soft radiation processes, providing more realistic events.

We deliberately do not include thermal background contributions in our analysis to enable direct comparison between pp and PbPb collision systems, isolating jet modification signatures induced by the medium without thermal complexity. Additionally, our comparative methodology includes a systematic evaluation of how the two different medium descriptions affect machine learning classification performance by using in-domain and cross-domain evaluations. The former compares the results of models trained and tested within the same medium, and the latter compares models trained in one medium and tested in the other.

\section{\label{sec:observables}Observables}
Particles are clustered into jets via the anti-k$_T$ algorithm \cite{cacciari2008anti} with a distance parameter $R = 0.4$. After the initial jet reconstruction, we apply the Cambridge/Aachen algorithm \cite{dokshitzer1997better} to recluster the jet constituents, building an angular-ordered tree structure that enables a step-by-step analysis of the internal splittings of the jet. The Soft Drop declustering \cite{larkoski2014soft} procedure is applied, continuously removing soft branches and keeping the splittings that satisfy the soft drop condition
\begin{equation}\label{eq:softdropcondition}
    z_\mathrm{g} \equiv \frac{\mathrm{min}(p_{T1}, p_{T2})}{p_{T1} + p_{T2}} \geq z_{\mathrm{cut}}\left( \frac{\Delta R}{R_0} \right)^\beta,
\end{equation}
where $p_{T1}$ and $p_{T2}$ are the transverse momenta of the two separated subjets, $\Delta R$ is the angular distance between them, and $R_0$ is the jet radius parameter, fixed in this work to $R_0 = 0.4$, as previously mentioned. The parameters $z_{\mathrm{cut}}$ and $\beta$ set the minimum allowed momentum fraction between the two branches in a splitting, and the angular dependence of the grooming condition, respectively.  

\begin{figure}
    \centering
    \includegraphics[width=1.0\linewidth]{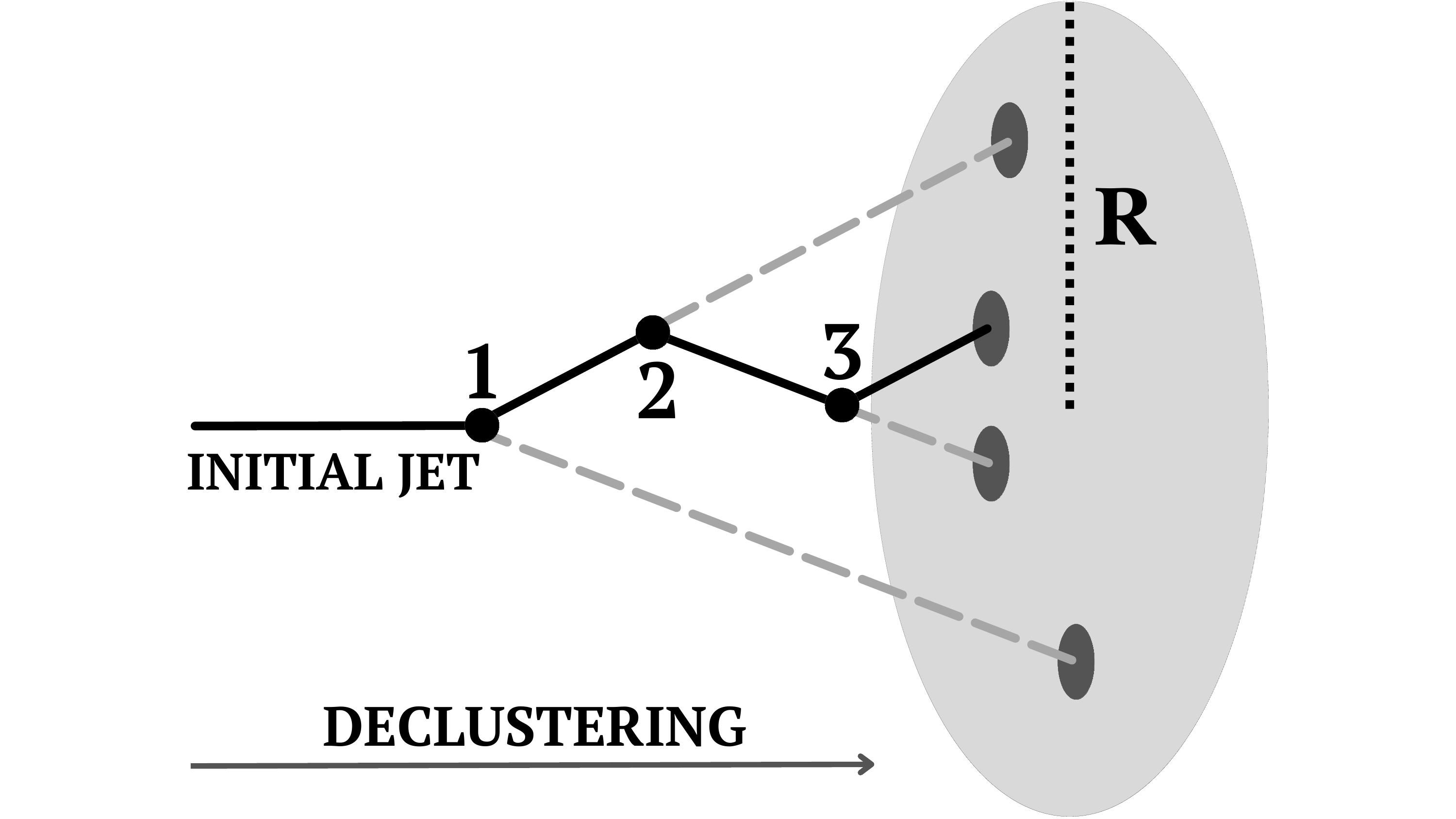}
    \caption{\justifying Schematic representation of the jet declustering history. Starting from the initial jet, the declustering procedure sequentially unfolds the branching structure through successive steps, labeled 1, 2, and 3, until reaching the final constituents inside the jet cone of radius \(R\). The solid black line highlights one possible declustering path, while the dashed gray lines indicate discarded branches of the jet tree by the Soft Drop procedure. Image inspired by \cite{CMS:2024:LundPlane}.}
    \label{fig:jet_soft_drop}
\end{figure}

In this study, we set the free parameters to $z_{\mathrm{cut}} = 0.1$ \cite{larkoski2014soft} and $\beta = 0$ \cite{larkoski2015sudakov}. This choice is motivated by the fact that this configuration provides a theoretically clean grooming procedure. In particular, $\beta = 0$ removes the angular dependence from the grooming condition, making the algorithm sensitive only to the momentum sharing between subjets, while $z_{\mathrm{cut}} = 0.1$ suppresses soft contaminating radiation without significantly affecting the hard core of the jet.

The Soft Drop algorithm works by removing the softer branch in each splitting until it encounters a hard splitting that satisfies the condition given by Eq. \eqref{eq:softdropcondition}. Then, the declustering proceeds along the harder subjet, as illustrated in Fig. \ref{fig:jet_soft_drop}.

In each splitting of the jet, we can compute substructure variables, such as the momentum fraction $z$, the angular distance $\Delta R$, perpendicular momentum $k_\perp$, and invariant mass $m_{\mathrm{inv}}$. These variables were chosen since they encapsulate key features of the jets, such as the shared energy, the opening angle between two splittings, the quantity of transverse energy from one branch to another, and the combination of the four-momenta of the two splittings. They can be calculated from the corresponding pair $(i, j)$ of declustered subjets
\begin{equation}
\begin{aligned}
    z =& \frac{\min(p_{T,i},\, p_{T,j})}{p_{T,i} + p_{T,j}}, \\
    \Delta R =& \sqrt{(\phi_i - \phi_j)^2 + (\eta_i - \eta_j)^2}, \\
    k_\perp =& \min(p_{T,i},\, p_{T,j}) \cdot \Delta R, \\
    m_{\mathrm{inv}} =& \sqrt{(E_i + E_j)^2 - (\mathbf{p}_i + \mathbf{p}_j)^2},
\end{aligned}
\end{equation}
where $i,j$ denotes the subjets at declustering step $t$. We store these variables in a vector $x_t$, so we have
\begin{equation}
    \mathbf{x}_t = [z,\, \Delta R,\, k_\perp,\, m_{\mathrm{inv}}],
\end{equation}
for each step $t$. Therefore, each jet can be represented as a sequence $\mathbf{x}$ of vectors $x_t$ that contains the information of its dynamical evolution. For a jet with $N$ declustering steps, we write
\begin{equation}\label{eq:vector_of_vectors}
    \mathbf{x} = [\mathbf{x}_1, \mathbf{x}_2, \dots, \mathbf{x}_t, \dots, \mathbf{x}_N].
\end{equation}

For each jet, a vector $\mathbf{x}$ in Eq. \eqref{eq:vector_of_vectors} can be associated as the declustering history, and we refer to this as sequential. However, one can associate each jet only to the first grooming step, $\mathbf{x}_1$, which we denote as non-sequential or static data.

The vector presented in Eq. \eqref{eq:vector_of_vectors} includes important information about the internal structure of jets, highlighting their interaction with the medium. In this manuscript, we investigate how this information evolves during the jet’s progression and examine the influence of different media, the simplified one used in \textsc{Jewel} and the more elaborated one given by v-USPhydro. Our objective is to understand how these substructures help in classifying quenching phenomena. For this purpose, we utilize Long Short-Term Memory (LSTM) and Transformer neural networks, as they are well-suited for working with sequential data. This sequential approach is strongly inspired by the findings in Ref. \cite{liu2023identifying}.

We also investigate the performance of models trained on static representations of the jets, which are used as input to machine learning models such as Random Forests and Multilayer Perceptrons, since these models do not explicitly capture the temporal dynamics of the jet evolution. As such, they serve as baselines for determining the discriminative power of global jet substructure patterns.

\section{\label{sec:mlmodels}Machine Learning Models}

The investigation in this manuscript employs a strategically structured set of machine learning methodologies, encompassing both traditional ensemble techniques and sophisticated sequential architectures, to systematically probe the discriminative signatures of jet quenching phenomena.

We present an architectural progression from non-sequential (Random Forest and MLP) to sequential (LSTM, LSTM+Attention and Transformer) models that reflects increasing sophistication in temporal feature extraction capabilities. Static (non-sequential) models evaluate discrete snapshots of jet substructure, while sequential architectures integrate complete evolutionary histories. This methodological hierarchy enables a systematic evaluation of whether temporal dynamics provide essential discriminative information beyond static observables.

We also employ cross-domain evaluation to ensure generalization capabilities across different medium descriptions, thereby validating the physical relevance of learned feature representations rather than specific artifacts of simulations.

The hyperparameters of each model are specified in Appendix \ref{app:hyperparameters}.
\subsection{Random Forest}
Random Decision Forests \cite{breiman2001random}, also known as Random Forests, are an ensemble learning method \cite{zhou2025ensemble, dietterich2000ensemble}, generally used for classification and regression problems. To make its predictions, this model combines the outputs of multiple decision trees \cite{breiman2017classification, quinlan1986induction} to reach a single result. 

The Random Forest algorithm is an extension of the bagging method, as it combines both bootstrap aggregation (bagging) \cite{breiman1996bagging} and random feature selection \cite{ho1998random} to grow an ensemble of uncorrelated forests of decision trees, improving generalization performance. 

This model was selected in this study as a baseline for comparison with subsequent neural network models, primarily because of its ability to capture global information about the features. The model's capacity to capture global feature interactions without temporal dependencies makes it particularly suited for evaluating the discriminative power of static jet substructure observables. Random Forest has been employed in HEP for resolving combinatorial ambiguities in dilepton $\bar{t}t$ event topologies \cite{alhazmi2022resolving}.

\subsection{Multilayer Perceptron}
Multilayer Perceptrons (MLPs) are a type of artificial neural network \cite{rosenblatt1958perceptron}, mathematically modeled to imitate the behavior of the human brain, consisting of multiple layers of neurons \cite{rumelhart1986learning}. A traditional MLP neuron utilizes nonlinear activation functions \cite{bishop2006pattern}, such as ReLU \cite{nair2010rectified}, which allows the model to capture more complex data patterns. This model has previously been employed in high-energy physics for real-time jet tagging \cite{sun2025fast}.

The performance of MLPs establishes a critical benchmark for assessing whether sequential modeling provides substantive advantages over static feature representations. Its strength lies in its capacity to model global interactions between input features without the need to rely on specific assumptions such as locality or sequential order. In this context, MLPs serve as a middle ground between traditional ensemble models and more advanced architectures, such as recurrent and attention-based networks.

\subsection{Long Short-Term Memory}
The Long Short-Term Memory (LSTM) is a type of Recurrent Neural Network (RNN) \cite{elman1990finding} designed to model sequential data and capture long-range dependencies over time. This model introduces a memory cell structure regulated by gating mechanisms (input, forget, and output gates) to enable selective information retention across extended sequences. These models have been successfully applied to a wide range of Natural Language Processing (NLP) tasks, such as machine translation \cite{hochreiter1997long, sutskever2014sequence}, and constituency parsing \cite{vinyals2015grammar}. They have also been employed in high-energy physics for top tagging at the LHC \cite{egan2017long}, and for jet quenching identification \cite{liu2023identifying}, which inspired part of this work.

%Given an input vector $\mathbf{x}_t$ at time step $t$, the previous output $\mathbf{h}_{t-1}$ and cell state $\mathbf{c}_{t-1}$, an LSTM with hidden size $k$ computes the next output $\mathbf{h}_t$ and cell state $\mathbf{c}_t$ as
%\begin{align}
%\mathbf{i}_t &= \sigma(\mathbf{W}^i \mathbf{H} + \mathbf{b}^i), \\
%\mathbf{f}_t &= \sigma(\mathbf{W}^f \mathbf{H} + \mathbf{b}^f), \\
%\mathbf{o}_t &= \sigma(\mathbf{W}^o \mathbf{H} + \mathbf{b}^o),  \\
%\mathbf{c}_t &= \mathbf{f}_t \odot \mathbf{c}_{t-1} + \mathbf{i}_t \odot \tanh(\mathbf{W}^c \mathbf{H} + \mathbf{b}^c), \\
%\mathbf{h}_t &= \mathbf{o}_t \odot \tanh(\mathbf{c}_t).
%\end{align}

%The indices $i$, $f$, $o$, and $c$ indicate the components associated with the input gate, forget gate, output gate, and cell update, respectively. Each gate has its own trainable weight matrix \( \mathbf{W}^{(\cdot)} \in \mathbb{R}^{2k \times k} \) and bias vector \( \mathbf{b}^{(\cdot)} \in \mathbb{R}^{k} \), where \( k \) is the size of the hidden state. The input vector at time \( t \), denoted by \( \mathbf{x}_t \), is concatenated with the previous hidden state \( \mathbf{h}_{t-1} \) to form the input \( \mathbf{H} \) to each gate, given by
%\begin{equation}
%    \mathbf{H} = \begin{bmatrix} \mathbf{x}_t \\ \mathbf{h}_{t-1} \end{bmatrix}.
%\end{equation}

%The function \( \sigma \) denotes a nonlinear activation function, typically a sigmoid function when applied to gates, while \( \tanh \) is the hyperbolic tangent function. The operator \( \odot \) represents the element-wise product between two vectors.

The forget gate controls the information that is discarded from the cell state, the input gate determines what must be stored, and the output gate decides what information is propagated to the next hidden state. These operations allow LSTMs to preserve relevant temporal information over sequences, making them an effective classifier for time-series tasks. This capability proves essential for modeling the progressive modifications occurring throughout jet evolution in dense media.

\subsubsection{Attention-Enhanced LSTM}
The attention mechanism \cite{bahdanau2014neural, luong2015effective} has transformed the approach to sequence modeling. It allows networks to focus on different parts of the input sequence dynamically when making predictions. Unlike traditional RNNs or LSTMs that fit everything into a single context vector, the attention mechanism takes a more nuanced approach by calculating a weighted sum of all hidden states. The weights are learned based on the importance of each time step for the current prediction.

The integration of an attention mechanism enhances the capability of the model to focus on relevant information within the input sequence after it has been processed. Specifically, for each output prediction generated by the LSTM, the attention layer assigns varying weights to each hidden state, pondering their contribution based on their relevance to the current prediction. This selective focus allows the model to identify and leverage critical substructures associated with the evolution of jets, resulting in greater interpretability and performance metrics. 

%This hybrid approach allows retrospective focus assignment across the complete grooming sequence. The attention weights offer direct insights into which evolutionary stages exhibit maximal discriminative power for quenching identification. The inclusion of attention mechanisms addresses the interpretability challenges inherent in deep models, providing direct visualization of temporal focus patterns. This capability may allow establishing connections between machine learning predictions and underlying physical mechanisms that govern jet-medium interactions.

Long Short-Term Memory coupled with attention has been used in physics, for instance, in accelerated prediction of charge density at the onset condition of DC corona discharge \cite{yi2022lstm}. To the best of our knowledge, studies explicitly exploring attention-enhanced architectures for quenched-jet classification are still limited.

A full mathematical description of this mechanism is beyond the scope of this work. For more details, see \cite{rocktaschel2015reasoning}.

\subsection{Transformer}

The Transformer architecture, introduced by Vaswani et al. \cite{vaswani2017attention}, is a deep learning model designed to manage sequential data. This architecture utilizes attention mechanism to effectively compute pairwise interactions between elements in an input sequence. This architecture has been applied in jet physics for tagging \cite{qu2022particle} and for quark-gluon discrimination and top tagging with a dual attention transformer \cite{he2023quark}, and, more recently, for the identification of jet-quenching \cite{CrispimRomao:2025yiq}.

Transformer Architecture represents the most sophisticated approach, and it overcomes the sequential processing limitations of recurrent models, enabling parallel computation of complex feature relationships across the entire jet substructure evolution. The multi-head attention framework \cite{vaswani2017attention} facilitates the capture of both local splitting characteristics and global evolutionary patterns.

%In this framework, given an input sequence composed of vectors denoted as \(\{\mathbf{x}_1, \mathbf{x}_2, \dots, \mathbf{x}_T\}\), each vector is transformed into three separate spaces named queries (\(\mathbf{Q}\)), keys (\(\mathbf{K}\)), and values (\(\mathbf{V}\)). 

%The mathematical expression of attention is 
%\begin{equation}
%    \text{Attention}(\mathbf{Q}, \mathbf{K}, \mathbf{V}) = \text{softmax}\left(\frac{\mathbf{Q} \mathbf{K}^\top}{\sqrt{d_k}}\right)\mathbf{V},
%\end{equation}
%where \(d_k\) represents the dimensionality of the key vectors and serves to scale the dot-product, preventing excessively large values that could affect the stability of the softmax function.

%Once this attention output is computed, it is passed through feedforward layers for further processing. Additionally, normalization techniques are applied, and these layers are often stacked multiple times to construct a deep encoder or decoder architecture.

Whereas LSTM with attention employs sequential processing by examining jet substructure elements chronologically, where a timestep depends on prior states, the Transformer architecture implements parallel self-attention, allowing simultaneous pairwise interactions across all grooming steps. 

\section{Results}

In this investigation, we use five different performance metrics to ensure rigorous quantitative evaluation across different operational conditions. \emph{Accuracy} measures the proportion of correctly classified jets relative to the total sample size. \emph{Area Under the Curve} (AUC) evaluates the discriminative capacity of the model across all classification thresholds, which is particularly valuable for imbalanced datasets. \emph{Precision} quantifies false positive rates by measuring the specificity of positive predictions and calculating the fraction of correctly identified quenched jets among all jets classified as quenched. \emph{Recall} evaluates the sensitivity of quenching detection, determining the proportion of actual quenched jets successfully identified by the model, thus measuring false negative rates. \emph{F1-Score} provides a harmonized metric balancing precision and recall through their harmonic mean, offering a consolidated performance indicator that accounts for both classification errors simultaneously. Please refer to \cite{ferrer2022analysis} for more information about supervised models evaluation metrics.

Hyperparameter optimization \cite{bergstra2012random, bergstra2011algorithms} was conducted by training several models with different parameter combinations on the training dataset, followed by an evaluation of their performance using the validation set. The configuration that achieved the best results on the validation set was selected for further use. Once the model was properly tuned, a final evaluation was performed with the test dataset, which remained completely isolated throughout the training and hyperparameter optimization processes. This approach ensures an unbiased evaluation of the predictive capabilities of the model. It also minimizes overfitting and selection bias, as it prevents the model from memorizing patterns that are specific to the training or validation data.

All datasets (training, validation and test) were generated under the same conditions of \textsc{Jewel} Default and \textsc{Jewel}+v-USPhydro, and remained strictly separate for each scenario. As discussed in Section \ref{sec:observables}, our dataset can be categorized into two main types: static and sequential. The tree-based model, Random Forest, and the fully connected neural network, Multilayer Perceptron (MLP), are trained and evaluated on the static Soft Drop variables computed from the Monte Carlo simulations. Whereas the dynamical evolution of the observables is fed to the recurrent neural network, the LSTM, and to the attention-based model, the Transformer.

To compute uncertainties associated with the performance metrics, we employ the non-parametric \emph{bootstrap} technique \cite{efron1992bootstrap, efron1994introduction}. It is widely used in machine learning and statistical modeling for quantifying uncertainty. For more details, see Refs. \cite{davison1997bootstrap, hernan2020causal}.

We use machine learning methods in an individual jet analysis, but we must address some methodological limitations. First of all, the jet reconstruction is not performed on top of a fully simulated thermal heavy-ion background, as would be the case in realistic Pb--Pb events. Instead, our analysis relies on the recoil partons generated by \textsc{Jewel}, which encode the medium response associated with the jet--medium interaction, but do not reproduce the full soft environment of an experimental heavy-ion collision. %Therefore, although recoils provide a physically meaningful representation of part of the in-medium response, they cannot be interpreted as a complete surrogate for the thermal background and its event-by-event structure.

This distinction is important because, in realistic high-multiplicity environments, jet measurements are affected not only by medium-induced modifications of the hard shower, but also by contamination from uncorrelated soft particles, local background fluctuations, and subtraction ambiguities. Such effects are known to degrade the precision of subjet and grooming observables, especially on a jet-by-jet basis, where the performance of subtraction procedures is substantially less robust than for ensemble-averaged quantities \cite{apolinario2013analysis, mengel2024multiplicity, mengel2023interpretable, budhraja2024jet}. Since our setup does not include a fully thermal Pb--Pb underlying event, the present results should not be understood as a full assessment of experimental background sensitivity, but rather as a controlled study of quenched-jet classification in a simplified medium-response scenario.

The recent work presented in \cite{ArrudaGoncalves:2025wtb} has investigated the robustness of machine-learning-based quenched-jet classification under underlying-event contamination, highlighting the importance of controlled comparisons between pp and PbPb-like samples.

A second limitation follows from the intrinsic ambiguity of energy-loss measurements. Jets selected with a given reconstructed transverse momentum do not correspond to a unique physical history, as they may originate either from much harder partons that lost a substantial fraction of their energy outside the jet cone, or from moderately energetic partons that experienced only mild modification \cite{connors2018jet}. This selection bias becomes particularly relevant in steeply falling jet spectra, where strongly quenched jets are statistically suppressed \cite{falcao2024constraining, andres2025minimizing}. As a consequence, even when machine learning models identify differences between classes, part of the discrimination power may still reflect this kinematic and selection bias rather than a clean isolation of medium-induced substructure effects. In this sense, the classification task should be interpreted with caution, since low-\(p_T\) jets may resemble quenched jets not only because of genuine in-medium modification, but also because of the interplay between recoil contamination, jet energy loss, and the biased population selected after reconstruction \cite{liu2023identifying, budhraja2024jet}.

In light of this issue, a related recent study used Transformer architectures on low-level constituent information to quantify the fraction of vacuum-like jets in heavy-ion samples, including medium response and underlying-event contamination \cite{CrispimRomao:2025yiq}.

We fully acknowledge these limitations and interpret results cautiously. Our objective is not absolute jet-level classification, but exploring whether statistical patterns consistent with medium-induced modifications can be systematically learned and interpreted within current simulation and analysis method constraints. This approach recognizes the inherent methodological boundaries while extracting physical insights from data.

For machine learning applications in jet quenching under realistic experimental conditions, including thermal background, detector effects and pileup, see \cite{qureshi2024model}.

\subsection{\label{sec:static_results}Static Data Models}
\subsubsection{Performance Metrics}
Tables \ref{tab:rf_metrics} and \ref{tab:mlp_metrics} show the evaluation metrics measured for the Random Forest and MLP, respectively.  The RF models exhibit a consistent performance across all $p_T$ intervals and for both the Default and v-USPhydro datasets. In particular, AUC scores remain above 0.85 in all cases, with a precision of around 0.5, indicating significant false positive classification.

\begin{table}[htb]
\centering
\caption{
Performance metrics for the Random Forest classifier across $p_T$ ranges and media.
}
\begin{ruledtabular}
\begin{tabular}{lccc}
$p_T$ [GeV] & Metric & Default & v-USPhydro \\
\hline
\multirow{5}{*}{40--60}
    & Accuracy   & 0.7683 (0.0006) & 0.7831 (0.0006) \\
    & AUC        & 0.8588 (0.0005) & 0.8737 (0.0005) \\
    & Precision  & 0.5295 (0.0011) & 0.6318 (0.0010) \\
    & Recall     & 0.8256 (0.0010) & 0.8322 (0.0008) \\
    & F1-Score   & 0.6452 (0.0010) & 0.7183 (0.0008) \\
\hline
\multirow{5}{*}{80--250}
    & Accuracy   & 0.7610 (0.0004) & 0.8065 (0.0003) \\
    & AUC        & 0.8601 (0.0003) & 0.8866 (0.0003) \\
    & Precision  & 0.5105 (0.0007) & 0.6646 (0.0006) \\
    & Recall     & 0.8171 (0.0006) & 0.8272 (0.0005) \\
    & F1-Score   & 0.6284 (0.0006) & 0.7370 (0.0005) \\
\hline
\multirow{5}{*}{200--400}
    & Accuracy   & 0.7776 (0.0006) & 0.8173 (0.0005) \\
    & AUC        & 0.8636 (0.0006) & 0.8896 (0.0004) \\
    & Precision  & 0.5253 (0.0011) & 0.6722 (0.0010) \\
    & Recall     & 0.7886 (0.0011) & 0.8285 (0.0008) \\
    & F1-Score   & 0.6306 (0.0010) & 0.7422 (0.0007) \\
\end{tabular}
\end{ruledtabular}
\label{tab:rf_metrics}
\end{table}

Although Multilayer Perceptrons (MLPs) are often seen as more flexible and capable due to their ability to represent complex, non-linear relationships, the findings indicate that the Random Forest model may actually perform better in this context. The MLP achieves slightly lower performance, with AUC values around 0.80 to 0.84. This performance hierarchy suggests that ensemble tree-based methods more effectively capture the jet substructure signatures relevant for quenching classification within the static feature space.

A critical limitation emerges in the precision-recall characteristics of both architectures. While recall values remain consistently high (0.75--0.85), precision scores are substantially lower (0.45--0.7), indicating systematic over-prediction of jet quenching events. This precision-recall imbalance suggests that static substructure observables contain overlapping signatures between quenched and unquenched jets, preventing clear discrimination based solely on static jet features.

From Tables \ref{tab:rf_metrics} and \ref{tab:mlp_metrics}, one can notice that the v-USPhydro medium consistently yields enhanced classification metrics compared to the Default \textsc{Jewel} configuration. For the Random Forest models, precision improves from 0.51--0.53 in the Default setup to 0.63--0.67 in v-USPhydro across all $p_T$ intervals. This improvement indicates that v-USPhydro generates more pronounced substructure modifications during jet-medium interactions.

\begin{table}[htb]
\centering
\caption{
Performance metrics for the MLP classifier across $p_T$ ranges and scenarios.
}
\begin{ruledtabular}
\begin{tabular}{lccc}
$p_T$ [GeV] & Metric & Default & v-USPhydro \\
\hline
\multirow{5}{*}{40--60}
    & Accuracy   & 0.7344 (0.0006) & 0.7095 (0.0006) \\
    & AUC        & 0.8410 (0.0006) & 0.7969 (0.0006) \\
    & Precision  & 0.4881 (0.0010) & 0.5452 (0.0010) \\
    & Recall     & 0.8342 (0.0010) & 0.7586 (0.0009) \\
    & F1-Score   & 0.6159 (0.0009) & 0.6344 (0.0008) \\
\hline
\multirow{5}{*}{80--250}
    & Accuracy   & 0.7286 (0.0004) & 0.7482 (0.0003) \\
    & AUC        & 0.8387 (0.0004) & 0.8113 (0.0004) \\
    & Precision  & 0.4717 (0.0006) & 0.6017 (0.0006) \\
    & Recall     & 0.8124 (0.0007) & 0.6850 (0.0007) \\
    & F1-Score   & 0.5969 (0.0006) & 0.6407 (0.0005) \\
\hline
\multirow{5}{*}{200--400}
    & Accuracy   & 0.7776 (0.0006) & 0.7607 (0.0005) \\
    & AUC        & 0.8438 (0.0006) & 0.8129 (0.0006) \\
    & Precision  & 0.5280 (0.0012) & 0.6102 (0.0010) \\
    & Recall     & 0.7179 (0.0012) & 0.6825 (0.0011) \\
    & F1-Score   & 0.6085 (0.0010) & 0.6443 (0.0009) \\
\end{tabular}
\end{ruledtabular}
\label{tab:mlp_metrics}
\end{table}

\subsubsection{\label{subsubsec:cross_domain}Cross-domain Validation}
To further examine the reliability of these models and eliminate concerns about the dependence on superficial patterns, we conducted cross-domain evaluations (see Tables~\ref{tab:rf_crossdomain} and~\ref{tab:mlp_crossdomain}). We trained models on jets originating in one medium and tested them on jets from the other medium, i.e., the classifier trained on the \textsc{Jewel} Default results is evaluated on data from \textsc{Jewel} + v-USPhydro and vice-versa.

\begin{table}[htb]
\centering
\caption{Random Forest cross-domain performance. Models trained on Default and tested on v-USPhydro, and vice-versa, for different $p_T$ intervals.}
\begin{ruledtabular}
\begin{tabular}{lccc}
$p_T$ [GeV] & Metric & Def$\rightarrow$v-USP & v-USP$\rightarrow$Def \\
\hline
\multirow{5}{*}{40--60}
    & Accuracy   & 0.7054 (0.0006) & 0.7571 (0.0006) \\
    & AUC        & 0.7593 (0.0006) & 0.8281 (0.0006) \\
    & Precision  & 0.5504 (0.0011) & 0.5165 (0.0012) \\
    & Recall     & 0.6186 (0.0011) & 0.7523 (0.0011) \\
    & F1-Score   & 0.5825 (0.0009) & 0.6125 (0.0010) \\
\hline
\multirow{5}{*}{80--250}
    & Accuracy   & 0.7263 (0.0004) & 0.7766 (0.0004) \\
    & AUC        & 0.7889 (0.0004) & 0.8393 (0.0004) \\
    & Precision  & 0.5676 (0.0006) & 0.5361 (0.0008) \\
    & Recall     & 0.6929 (0.0006) & 0.7160 (0.0008) \\
    & F1-Score   & 0.6240 (0.0005) & 0.6131 (0.0007) \\
\hline
\multirow{5}{*}{200--400}
    & Accuracy   & 0.7421 (0.0005) & 0.7880 (0.0006) \\
    & AUC        & 0.7958 (0.0006) & 0.8424 (0.0006) \\
    & Precision  & 0.5810 (0.0010) & 0.5457 (0.0012) \\
    & Recall     & 0.6733 (0.0010) & 0.7123 (0.0012) \\
    & F1-Score   & 0.6237 (0.0008) & 0.6180 (0.0010) \\
\end{tabular}
\end{ruledtabular}
\label{tab:rf_crossdomain}
\end{table}

Cross-domain validation reveals significant performance degradation when models trained on one medium are evaluated on jets from the other medium description. Random Forest AUC scores decrease to the 0.76--0.84 range, while MLP performance drops to 0.73--0.82. Notably, there is an asymmetric generalization pattern that emerges, since models trained on v-USPhydro and tested on Default configurations (v-USPhydro $\rightarrow$ Default) outperform the reverse direction (Default $\rightarrow$ v-USPhydro), as evidenced by the increase in accuracy, AUC and recall scores. However, precision is still about $50\%$, so the models continue predicting a significant amount of false positives.

This asymmetry suggests that v-USPhydro creates more generalizable jet modification patterns that partially transfer to Default medium descriptions. Conversely, models trained on Default fail to capture the modification signatures present in v-USPhydro, indicating that the latter incorporates additional outcomes not present in the former.

The analysis of false positives (FPs) across media and cross-domain scenarios show that in the in-domain context, the Default medium demonstrates a consistently higher rate of false positives than the v-USPhydro medium, as evidenced by its lower precision values (0.47–0.53 compared to 0.54–0.67). %This could imply that the Default medium generates jet patterns that are inherently more ambiguous and susceptible to being misidentified as signatures of quenching. In contrast, the quenching features in the hydrodynamical medium are clearer and less likely to be confused.

The situation becomes even more pronounced in a cross-domain setting. When models trained on v-USPhydro are tested on Default data (v-USPhydro → Default), there is a notable increase in recall, but this comes at the cost of precision, resulting in a higher occurrence of false positives. This observation indicates that models trained in the more realistic hydrodynamical medium may tend to over-interpret structures in the Default samples as indications of quenching. 

Conversely, when training is performed on the Default medium and testing on v-USPhydro (Default \(\rightarrow\) v-USPhydro), the models tend to exhibit higher precision but lower recall. This behavior indicates a more conservative classification regime, in which quenched jets are identified less often, but with greater confidence. Such a pattern suggests that the v-USPhydro medium imprints a broader and more diverse range of modifications on the final observables. As a consequence, a model trained exclusively on the Default scenario may learn a decision boundary that is too restrictive to capture the full variety of quenched jets present in the v-USPhydro case, leading to missed positive instances and therefore reduced recall.

This trade-off between precision and recall underscores a fundamental disparity in the characteristics of the two media. Training on v-USPhydro enhances sensitivity to subtle structures, which can lead to a greater rate of false positives in the v-USPhydro $\rightarrow$ Default context. Conversely, training on Default confines the classifier to more robust signatures, thereby elevating precision at the risk of overlooking numerous true quenched jets in the hydrodynamical context. These findings emphasize that the choice of training medium plays a critical role in determining the equilibrium between overestimation and underestimation of quenching phenomena, reinforcing the necessity for models capable of effectively disentangling universal physical features from model-specific artifacts.

\begin{table}[htb]
\centering
\caption{MLP cross-domain performance. Models trained on Default and tested on v-USPhydro, and vice-versa, for different $p_T$ intervals.}
\begin{ruledtabular}
\begin{tabular}{lccc}
$p_T$ [GeV] & Metric & Def$\rightarrow$v-USP & v-USP$\rightarrow$Def \\
\hline
\multirow{5}{*}{40--60}
    & Accuracy   & 0.6512 (0.0006) & 0.7172 (0.0006) \\
    & AUC        & 0.7311 (0.0007) & 0.8178 (0.0006) \\
    & Precision  & 0.4784 (0.0011) & 0.4687 (0.0010) \\
    & Recall     & 0.5525 (0.0011) & 0.8110 (0.0011) \\
    & F1-Score   & 0.5128 (0.0010) & 0.5941 (0.0009) \\
\hline
\multirow{5}{*}{80--250}
    & Accuracy   & 0.7034 (0.0003) & 0.7521 (0.0004) \\
    & AUC        & 0.7732 (0.0004) & 0.8188 (0.0004) \\
    & Precision  & 0.5360 (0.0006) & 0.4991 (0.0007) \\
    & Recall     & 0.7083 (0.0006) & 0.6703 (0.0008) \\
    & F1-Score   & 0.6102 (0.0005) & 0.5721 (0.0007) \\
\hline
\multirow{5}{*}{200--400}
    & Accuracy   & 0.7399 (0.0005) & 0.7710 (0.0006) \\
    & AUC        & 0.7754 (0.0006) & 0.8271 (0.0006) \\
    & Precision  & 0.5856 (0.0011) & 0.5183 (0.0012) \\
    & Recall     & 0.6180 (0.0011) & 0.6884 (0.0012) \\
    & F1-Score   & 0.6014 (0.0009) & 0.5913 (0.0010) \\
\end{tabular}
\end{ruledtabular}
\label{tab:mlp_crossdomain}
\end{table}

Another important trend that is observed concerns the recall values across different transverse momentum ($p_T$) ranges in the MLP model. In the in-domain setting, recall is consistently lower for the v-USPhydro model compared to the Default model, particularly at lower $p_T$ values. This discrepancy suggests that when both models are trained and tested within the hydrodynamical medium, the v-USPhydro model fails to detect a larger fraction of quenched jets. This phenomenon aligns with the expectation that the quenching effects represented in v-USPhydro are more subtle and continuous, making them inherently more challenging for models to identify. In contrast, the Default medium tends to induce stronger and more discrete modifications to the jets, which facilitates easier recognition by the classifiers.

Interestingly, as $p_T$ increases, the recall values for both media begin to converge, with the differences becoming minimal within the $200$–$400$ GeV range. This observation implies that, at higher energies, the quenching signatures that the models capture tend to become more uniform, displaying less sensitivity to the specific description of the underlying medium. 

Despite inferior in-domain performance, MLPs demonstrate superior cross-domain generalization compared to Random Forest models. This is evidenced by a smaller performance gap when evaluating across different domains for MLPs. Implementing more complex neural networks for sequential data will clarify if the better performance of the MLP in cross-domain scenarios reflects a general advantage over tree-based models, as the Random Forest, in capturing medium dependence.

Overall, these results imply that relying solely on static substructure features for jet quenching classification yields limited discriminative capability, as evidenced by high false positive rates across different model architectures and medium configurations. The observed dependence on medium, along with poor cross-domain generalization, underscores the need for more advanced feature representations to effectively encapsulate the physics involved in jet modifications during heavy-ion collisions.

\subsubsection{Importance Analysis}

To explore which input features most influence the predictions of the model, we perform a feature importance analysis based on SHAP (SHapley Additive exPlanations) values \cite{lundberg2017unified}. This method is grounded in cooperative game theory and provides model-agnostic attributions of feature relevance. Positive SHAP values indicate that a given feature contributed to the prediction of the positive class (1), while negative values reveal that the feature pushes the prediction toward class 0.

In this work, quenched jets correspond to the positive class, 1, while unquenched jets are typed as 0.

\begin{figure*}[t]
    \centering

    \includegraphics[width=0.75\textwidth]{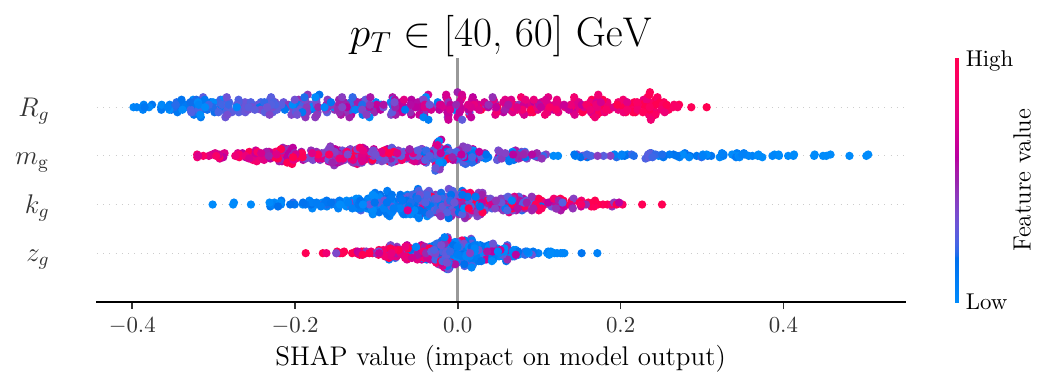}
    \caption*{(a) Default.}

    \vspace{0.3cm}

    \includegraphics[width=0.75\textwidth]{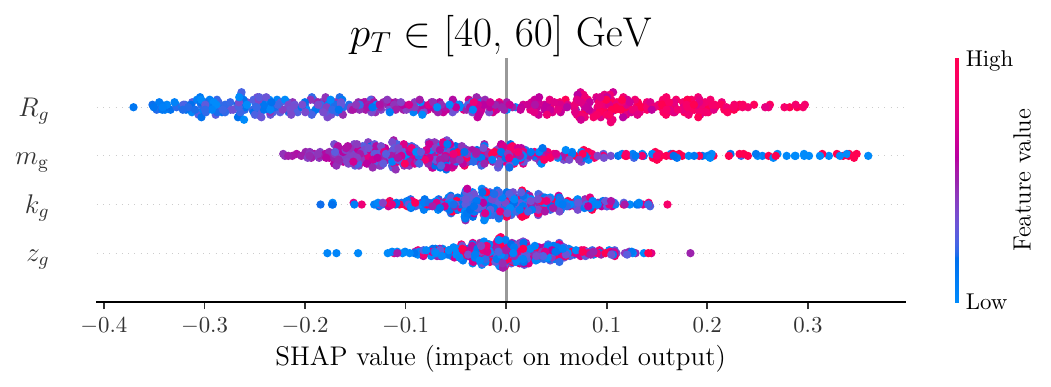}
    \caption*{(b) v-USPhydro.}

    \caption{
    SHAP values for $40 < p_T < 60$ GeV. \( R_g \) and \( m_g \) are the most important features in both media.
    }
    \label{fig:shap_40_60}
\end{figure*}

\begin{figure*}[t]
    \centering

    \includegraphics[width=0.65\textwidth]{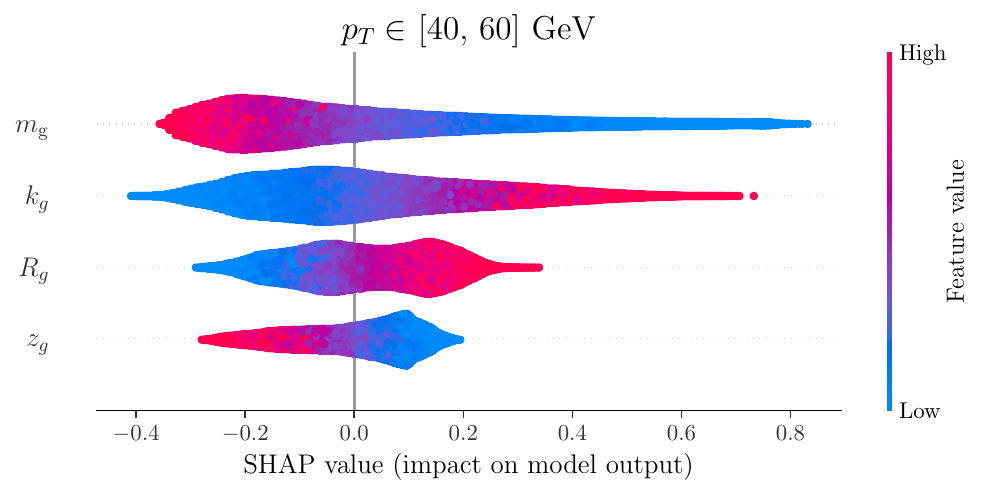}
    \caption*{(a) Default.}

    \vspace{0.3cm}

    \includegraphics[width=0.65\textwidth]{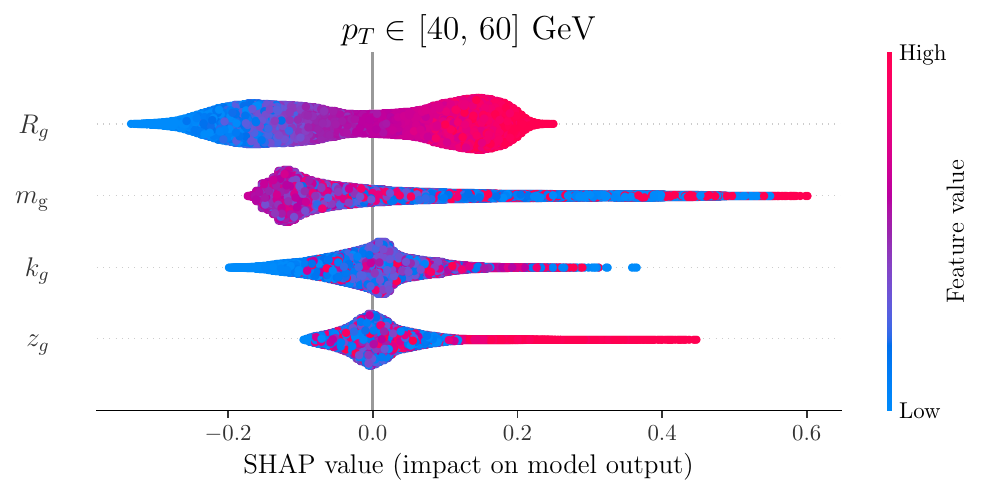}
    \caption*{(b) v-USPhydro.}

    \caption{
    Feature importance (SHAP values) for the MLP model in the $40 < p_T < 60$~GeV interval.
    }
    \label{fig:mlp_shap_40_60}
\end{figure*}

We display the results of the SHAP beeswarm plot in Figs. \ref{fig:shap_40_60} and \ref{fig:mlp_shap_40_60} for the Random Forest and MLP, respectively. The other two intervals of transverse momentum present the same tendency as the $p_T \in [40, 60]$ GeV results, and for brevity we do not exhibit them in this manuscript.

From Figures \ref{fig:shap_40_60} and \ref{fig:mlp_shap_40_60}, one can see that the groomed angular separation, $R_g$, and the groomed mass, $m_g$, are generally the two most relevant features in the decision-making of both models. Specifically, higher values of \(R_g\) are associated with the classification of jets as quenched, while greater values of \(m_g\) indicate that a jet is more likely to be classified as unquenched.

The tendencies observed manifest across all examined transverse momentum ranges, medium configurations, and the two ML approaches tested up to this point. The consistency of this pattern across different conditions suggests the presence of a physical signature.

%% 80_250 SHAP MLP
\begin{comment}
\begin{figure}[htbp]
    \centering
    % 80--250 GeV
    \begin{subfigure}{0.48\textwidth}
        \centering
        \includegraphics[width=\linewidth]{figures/summary_plot_MLP_default_80_250.pdf}
        \caption{Default}
    \end{subfigure}
    \hfill
    \begin{subfigure}{0.48\textwidth}
        \centering
        \includegraphics[width=\linewidth]{figures/summary_plot_MLP_vusp_80_250.pdf}
        \caption{v-USPhydro}
    \end{subfigure}
    \caption{Feature importance (SHAP values) for the MLP model in the $80 < p_T < 250$~GeV interval.}
    \label{fig:mlp_shap_80_250}
\end{figure}

\begin{figure}[htbp]
    \centering
    % 200--400 GeV
    \begin{subfigure}{0.48\textwidth}
        \centering
        \includegraphics[width=\linewidth]{figures/summary_plot_MLP_default_200_400.pdf}
        \caption{Default}
    \end{subfigure}
    \hfill
    \begin{subfigure}{0.48\textwidth}
        \centering
        \includegraphics[width=\linewidth]{figures/summary_plot_MLP_vusp_200_400.pdf}
        \caption{v-USPhydro}
    \end{subfigure}
    \caption{Feature importance (SHAP values) for the MLP model in the $200 < p_T < 400$~GeV interval.}
    \label{fig:mlp_shap_200_400}
\end{figure}
\end{comment}

%This characteristic of subjet separation initially appears to contradict recent experimental measurements from the ALICE Collaboration, which report jets becoming narrower after quenching in quark–gluon plasma distributions \cite{acharya2022measurement}. 

%\vspace{-0.3cm}

The research presented in \cite{milhano2018sensitivity}, which also utilized the \textsc{Jewel} framework, illustrates that recoil effects consistently alter jet substructure. These modifications result in an enhancement of angular separation, which aligns with the conclusions drawn from our analysis using SHAP values.

In the Default scenario, we observe a clearer separation in the SHAP values compared to the v-USPhydro scenario. This reinforces our point that the hydrodynamic medium introduces subtle variations in the quenching pattern. As a result, higher feature values become more intermixed with lower ones, although a visible trend still exists.

The difference in granularity observed in the SHAP beeswarm plots between the models is 
due to the specific explainers used. For the Random Forest, the TreeExplainer leverages the model's discrete tree structure, where multiple observations falling into the same leaf nodes receive identical SHAP values, resulting in distinct vertical clusters. In contrast, the MLP employs the KernelExplainer, which operates on continuous functions and gradients. This leads to a smoother, more organic distribution of points, reflecting the model's continuous nature.

\subsection{Sequential Data Models}
\subsubsection{Performance Metrics}

Tables \ref{tab:lstm_results} and \ref{tab:lstmatt_results} present the performance metrics for both LSTM and LSTM+Attention models, evaluated on the \textsc{Jewel} Default and v-USPhydro substructure datasets, across the three $p_T$ intervals that are considered in this study. In all cases, the values reported are averages over independent runs, with the error indicated as well. 

\begin{table}[htb]
\centering
\caption{Performance metrics for the LSTM model across $p_T$ intervals for Default and v-USPhydro datasets.}
\begin{ruledtabular}
\begin{tabular}{lccc}
$p_T$ [GeV] & Metric & Default & v-USPhydro \\
\hline
\multirow{5}{*}{40--60}
    & Accuracy   & 0.9695 (0.0002) & 0.9351 (0.0003) \\
    & AUC        & 0.9896 (0.0002) & 0.9727 (0.0002) \\
    & Precision  & 0.9578 (0.0005) & 0.9285 (0.0006) \\
    & Recall     & 0.9211 (0.0008) & 0.8720 (0.0008) \\
    & F1-Score   & 0.9391 (0.0005) & 0.8994 (0.0005) \\
\hline
\multirow{5}{*}{80--250}
    & Accuracy   & 0.9770 (0.0001) & 0.9447 (0.0002) \\
    & AUC        & 0.9917 (0.0001) & 0.9756 (0.0001) \\
    & Precision  & 0.9700 (0.0003) & 0.9339 (0.0003) \\
    & Recall     & 0.9371 (0.0004) & 0.8964 (0.0004) \\
    & F1-Score   & 0.9533 (0.0002) & 0.9148 (0.0003) \\
\hline
\multirow{5}{*}{200--400}
    & Accuracy   & 0.9778 (0.0002) & 0.9481 (0.0003) \\
    & AUC        & 0.9917 (0.0002) & 0.9757 (0.0002) \\
    & Precision  & 0.9530 (0.0006) & 0.9520 (0.0005) \\
    & Recall     & 0.9579 (0.0005) & 0.8858 (0.0007) \\
    & F1-Score   & 0.9554 (0.0004) & 0.9177 (0.0004) \\
\end{tabular}
\end{ruledtabular}
\label{tab:lstm_results}
\end{table}

Both the LSTM and LSTM+Attention models demonstrate exceptional classification performance. These models achieve accuracy and AUC scores consistently above 95\% across all ranges of transverse momentum. The incorporation of the attention mechanism does not increase the predictive capabilities, but rather, it maintains the same levels of performance. 

\begin{figure*}[htb]
    \centering
    % Primeira linha: duas figuras lado a lado
    \begin{subfigure}[b]{0.49\textwidth}
        \centering
        \includegraphics[width=\textwidth]{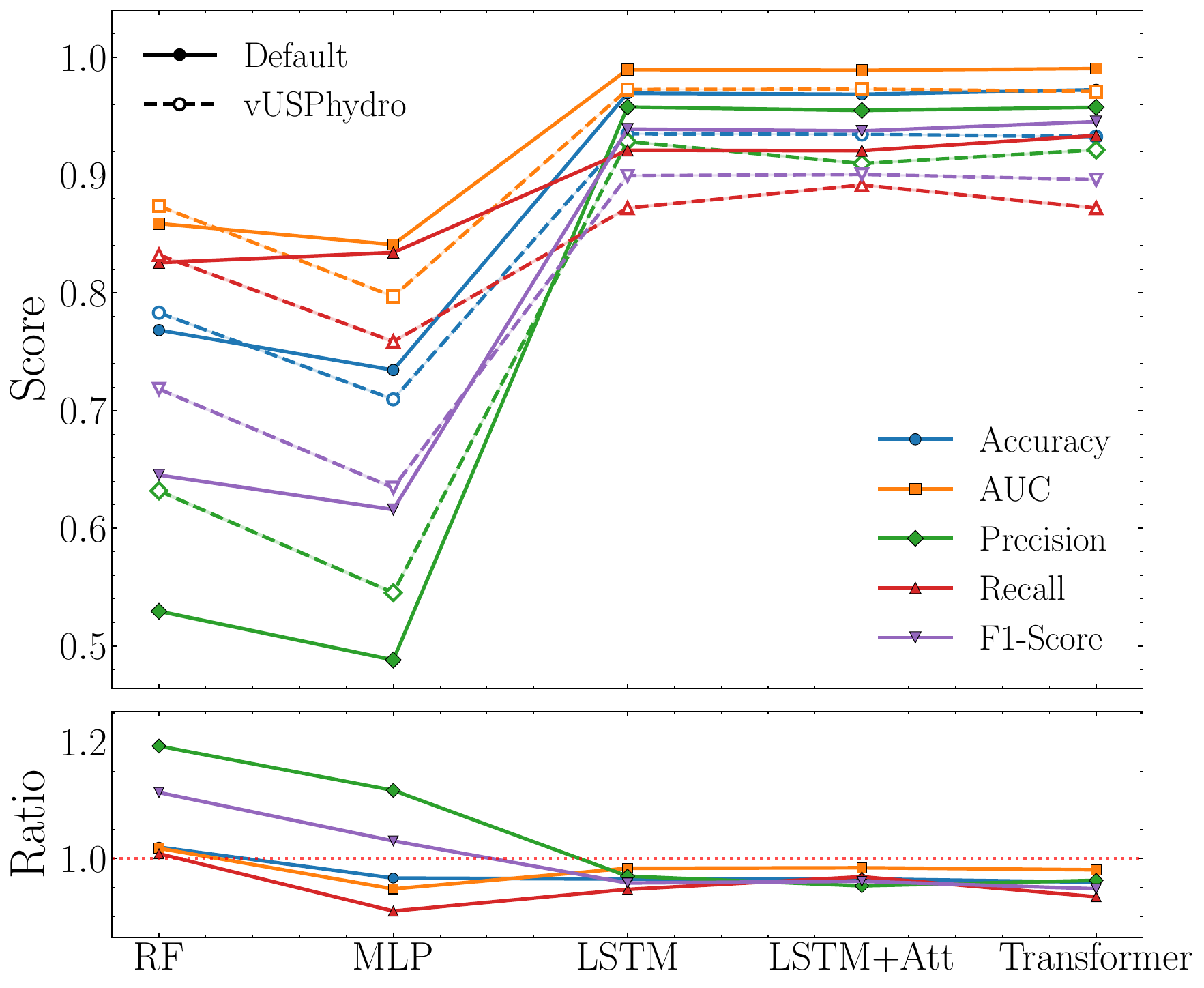}
        \caption{$p_T \in [40, 60]$ GeV}
        \label{fig:indomain_40_60}
    \end{subfigure}
    \hfill
    \begin{subfigure}[b]{0.49\textwidth}
        \centering
        \includegraphics[width=\textwidth]{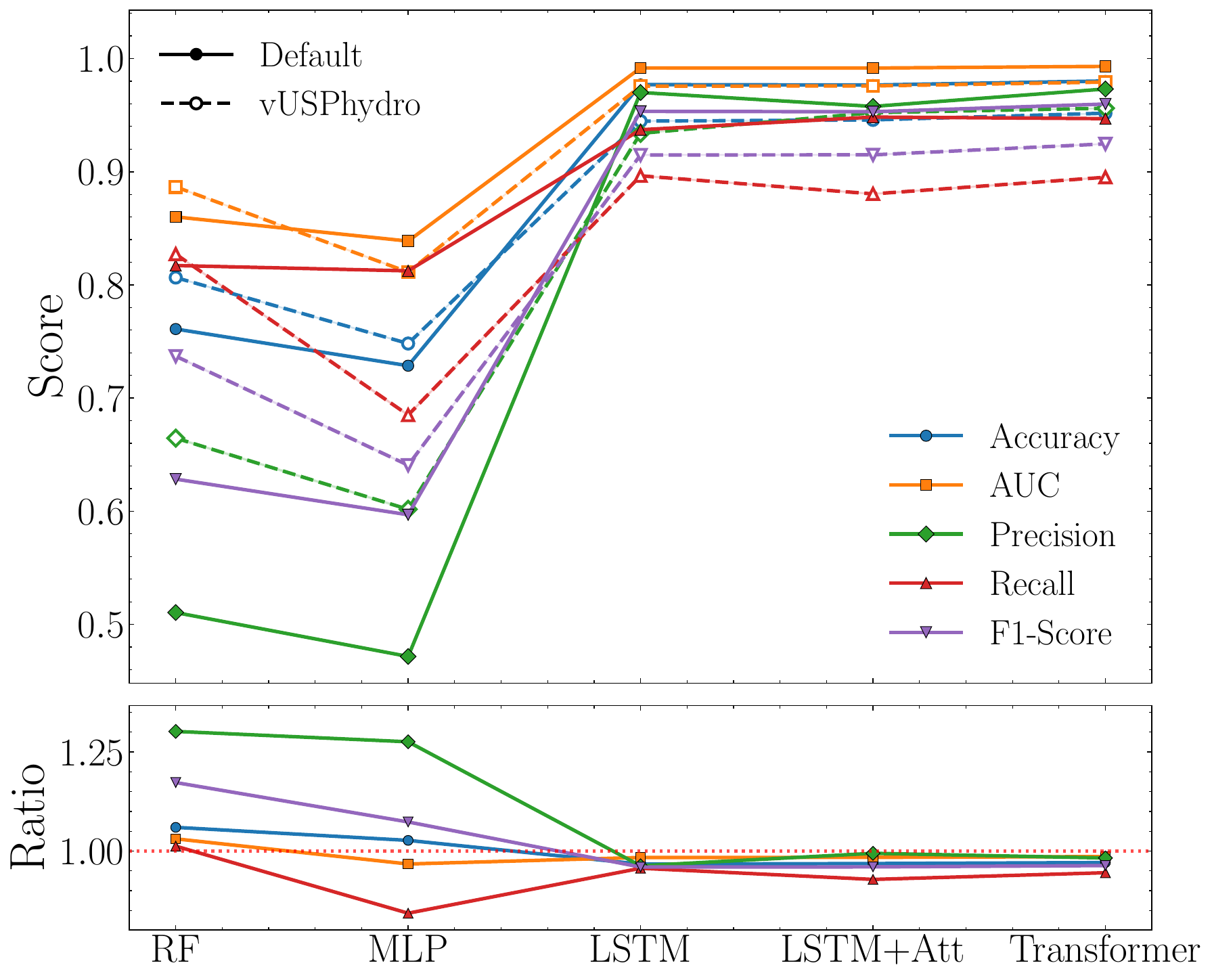}
        \caption{$p_T \in [80, 250]$ GeV}
        \label{fig:indomain_80_250}
    \end{subfigure}

    % Segunda linha: uma figura centralizada
    \vskip\baselineskip
    \begin{subfigure}[b]{0.5\textwidth}
        \centering
        \includegraphics[width=\textwidth]{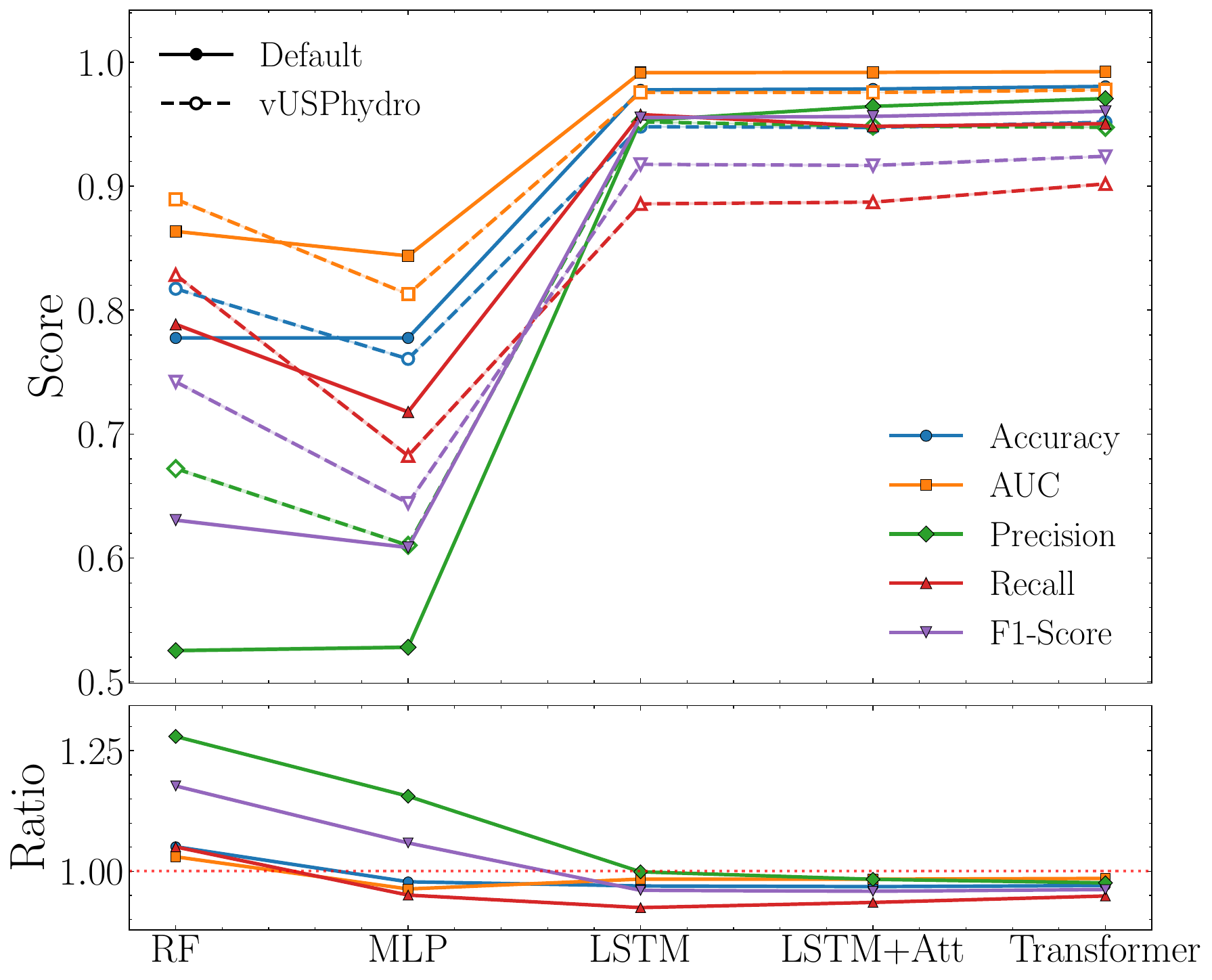}
        \caption{$p_T \in [200, 400]$ GeV}
        \label{fig:indomain_200_400}
    \end{subfigure}

    \caption{\justifying In-domain performance for each $p_T$ interval. Models are trained and tested on the same medium. All models perform well, with slightly lower performance observed in the v-USPhydro case. Here, the ratio is defined as the value of a given metric obtained with the v-USPhydro model divided by the corresponding value obtained with the Default model.}
    \label{fig:indomain_all}
\end{figure*}

In the in-domain scenario, both the LSTM and LSTM+Attention architectures exhibit a significant advantage when trained and evaluated on the Default medium compared to the v-USPhydro setup. Across all transverse momentum ($p_T$) intervals, the Default case demonstrates higher precision and recall simultaneously, culminating in superior $F_1$-scores. For example, within the $40$--$60~\text{GeV}$ range, the Default-trained LSTM achieves a precision of $0.95$ and a recall of $0.92$, while the corresponding values for the v-USPhydro configuration decrease to $0.92$ and $0.89$, respectively. This trend becomes even more pronounced at higher transverse momenta, where Default-trained models maintain recall values approaching $0.95$, in stark contrast to the systematically lower values seen in the v-USPhydro setup.

These results suggest that the Default setup offers a cleaner pattern for identifying quenching signatures. This allows the models to differentiate between quenched and unquenched jets with greater sensitivity and reliability. Conversely, the implementation of a realistic hydrodynamical medium, while more representative of actual heavy-ion collisions, introduces additional effects that obscure the quenching signals in the sequential representation, thereby diminishing the efficacy of the models in detecting quenched jets.

\begin{table}[htb]
\centering
\caption{Performance metrics for the LSTM+Attention model across $p_T$ intervals for Default and v-USPhydro datasets.}
\begin{ruledtabular}
\begin{tabular}{lccc}
$p_T$ [GeV] & Metric & Default & v-USPhydro \\
\hline
\multirow{5}{*}{40--60}
    & Accuracy   & 0.9686 (0.0002) & 0.9345 (0.0003) \\
    & AUC        & 0.9890 (0.0002) & 0.9730 (0.0002) \\
    & Precision  & 0.9549 (0.0006) & 0.9098 (0.0007) \\
    & Recall     & 0.9207 (0.0008) & 0.8916 (0.0007) \\
    & F1-Score   & 0.9375 (0.0005) & 0.9006 (0.0005) \\
\hline
\multirow{5}{*}{80--250}
    & Accuracy   & 0.9766 (0.0001) & 0.9458 (0.0002) \\
    & AUC        & 0.9916 (0.0001) & 0.9758 (0.0001) \\
    & Precision  & 0.9577 (0.0003) & 0.9522 (0.0003) \\
    & Recall     & 0.9483 (0.0004) & 0.8804 (0.0004) \\
    & F1-Score   & 0.9530 (0.0002) & 0.9149 (0.0003) \\
\hline
\multirow{5}{*}{200--400}
    & Accuracy   & 0.9785 (0.0002) & 0.9474 (0.0003) \\
    & AUC        & 0.9919 (0.0001) & 0.9757 (0.0002) \\
    & Precision  & 0.9645 (0.0005) & 0.9485 (0.0005) \\
    & Recall     & 0.9484 (0.0006) & 0.8872 (0.0007) \\
    & F1-Score   & 0.9564 (0.0004) & 0.9168 (0.0004) \\
\end{tabular}
\end{ruledtabular}
\label{tab:lstmatt_results}
\end{table}

The comparison between LSTM and LSTM+Attention models reveals that neither model consistently outperforms the other across all metrics and $p_T$ ranges. Both exhibit strong performance in the Default and v-USPhydro scenarios, although the results for the latter are generally less favorable. This suggests that while sequential models are more effective for jet classification in a simpler environment, they face greater challenges in a hydrodynamical medium, which introduces a more complex classification landscape. 

One possible explanation for the reason why the LSTM+Attention does not outperform the LSTM alone is that the jet substructure sequences used in our analysis are relatively short and may not exhibit strong long-range dependencies. In such cases, the internal gating mechanisms of the LSTM are already sufficient to capture the relevant temporal structure, and the added attention layer does not provide a significant advantage.

\begin{table}[htb]
\centering
\caption{Performance metrics for the Transformer model on Default and v-USPhydro datasets across different $p_T$ intervals. Values are presented as mean(error).}
\begin{ruledtabular}
\begin{tabular}{llcc}
$p_T$ [GeV] & Metric     & Default           & v-USPhydro            \\ \hline
\multirow{5}{*}{40--60}
           & Accuracy   & 0.9725 (0.0002) & 0.9327 (0.0003) \\
           & AUC        & 0.9905 (0.0002) & 0.9710 (0.0002) \\
           & Precision  & 0.9576 (0.0006) & 0.9214 (0.0006) \\
           & Recall     & 0.9336 (0.0007) & 0.8720 (0.0008) \\
           & F1-Score   & 0.9454 (0.0005) & 0.8960 (0.0005) \\ \hline
\multirow{5}{*}{80--250}
           & Accuracy   & 0.9802 (0.0001) & 0.9517 (0.0002) \\
           & AUC        & 0.9931 (0.0001) & 0.9792 (0.0001) \\
           & Precision  & 0.9731 (0.0003) & 0.9561 (0.0003) \\
           & Recall     & 0.9469 (0.0004) & 0.8951 (0.0004) \\
           & F1-Score   & 0.9598 (0.0002) & 0.9246 (0.0003) \\ \hline
\multirow{5}{*}{200--400}
           & Accuracy   & 0.9806 (0.0002) & 0.9517 (0.0003) \\
           & AUC        & 0.9925 (0.0001) & 0.9778 (0.0002) \\
           & Precision  & 0.9708 (0.0005) & 0.9476 (0.0005) \\
           & Recall     & 0.9505 (0.0006) & 0.9019 (0.0006) \\
           & F1-Score   & 0.9605 (0.0004) & 0.9242 (0.0004) \\
\end{tabular}
\end{ruledtabular}
\label{tab:transformer_results}
\end{table}

Table \ref{tab:transformer_results} shows the performance of the Transformer model for jet classification in the in-domain medium. As with the previous models, we evaluate accuracy, AUC, precision, recall, and F$_1$-score.

When we train and evaluate the model within the same conditions, the Transformer has excellent predictive power, as was the case for the LSTM. Accuracy and AUC score values are above $97\%$, indicating an almost perfect separation between quenched and unquenched samples. Also, there is a small decrease in recall for the v-USPhydro model, which can be due to the more complex nature and event-by-event variations in a realistic hydrodynamical medium, as explained for the LSTMs. The F$_1$-score displays significantly high values, indicating that the model does not require a trade-off between precision and recall to achieve strong results.

These results confirm that the Transformer effectively extracts and uses sequential information in jet substructure observables. The inner attention and self-attention mechanisms within the model enable it to dynamically focus on both local features and global patterns that emerge as the jet moves through the medium. The small differences in performance between the \textsc{Jewel} Default and v-USPhydro scenarios are not necessarily failures of the model but rather a reflection of the underlying dynamics that the realistic medium adds by changing the radiation phase space, which makes it harder to distinguish quenching phenomena, as discussed previously.

Figure \ref{fig:indomain_all} displays an intuitive visualization of the difference in performance metrics results for this work.

\subsubsection{Cross-domain Validation}

\begin{figure*}[htb]
    \centering
    % Primeira linha: duas figuras lado a lado
    \begin{subfigure}[b]{0.49\textwidth}
        \centering
        \includegraphics[width=\textwidth]{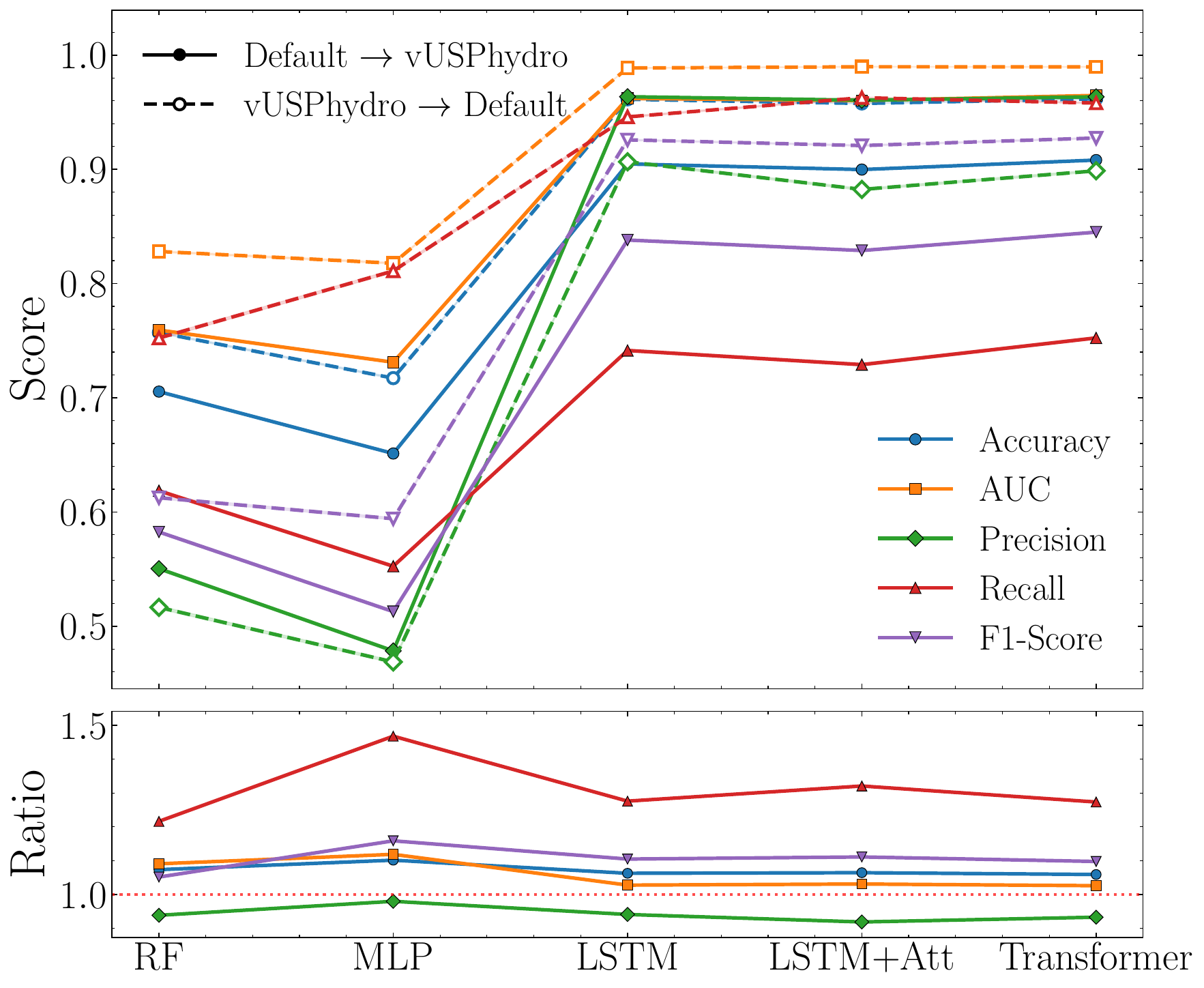}
        \caption{$p_T \in [40, 60]$ GeV}
        \label{fig:crossdomain_40_60}
    \end{subfigure}
    \hfill
    \begin{subfigure}[b]{0.49\textwidth}
        \centering
        \includegraphics[width=\textwidth]{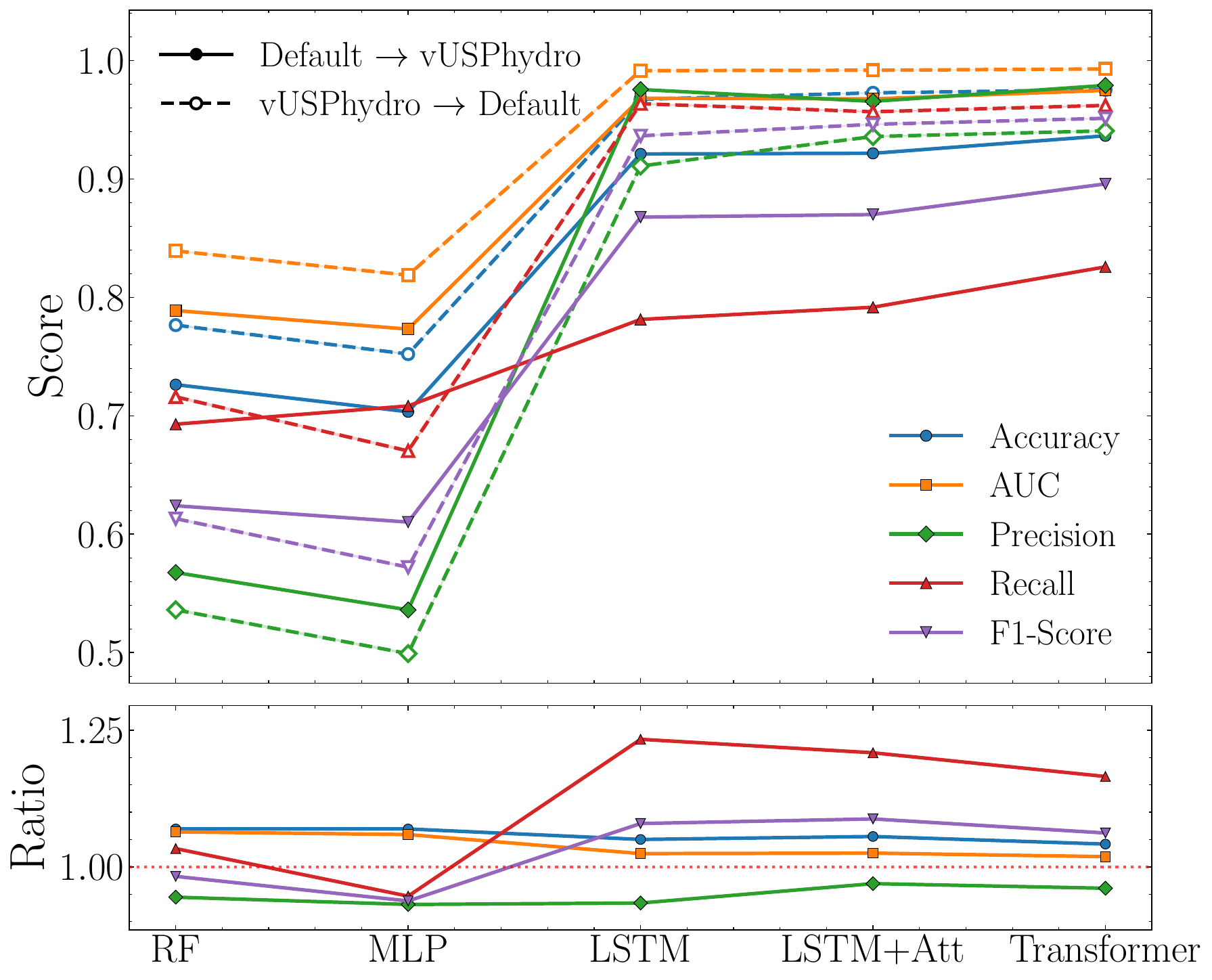}
        \caption{$p_T \in [80, 250]$ GeV}
        \label{fig:crossdomain_80_250}
    \end{subfigure}

    % Segunda linha: figura centralizada
    \vskip\baselineskip
    \begin{subfigure}[b]{0.5\textwidth}
        \centering
        \includegraphics[width=\textwidth]{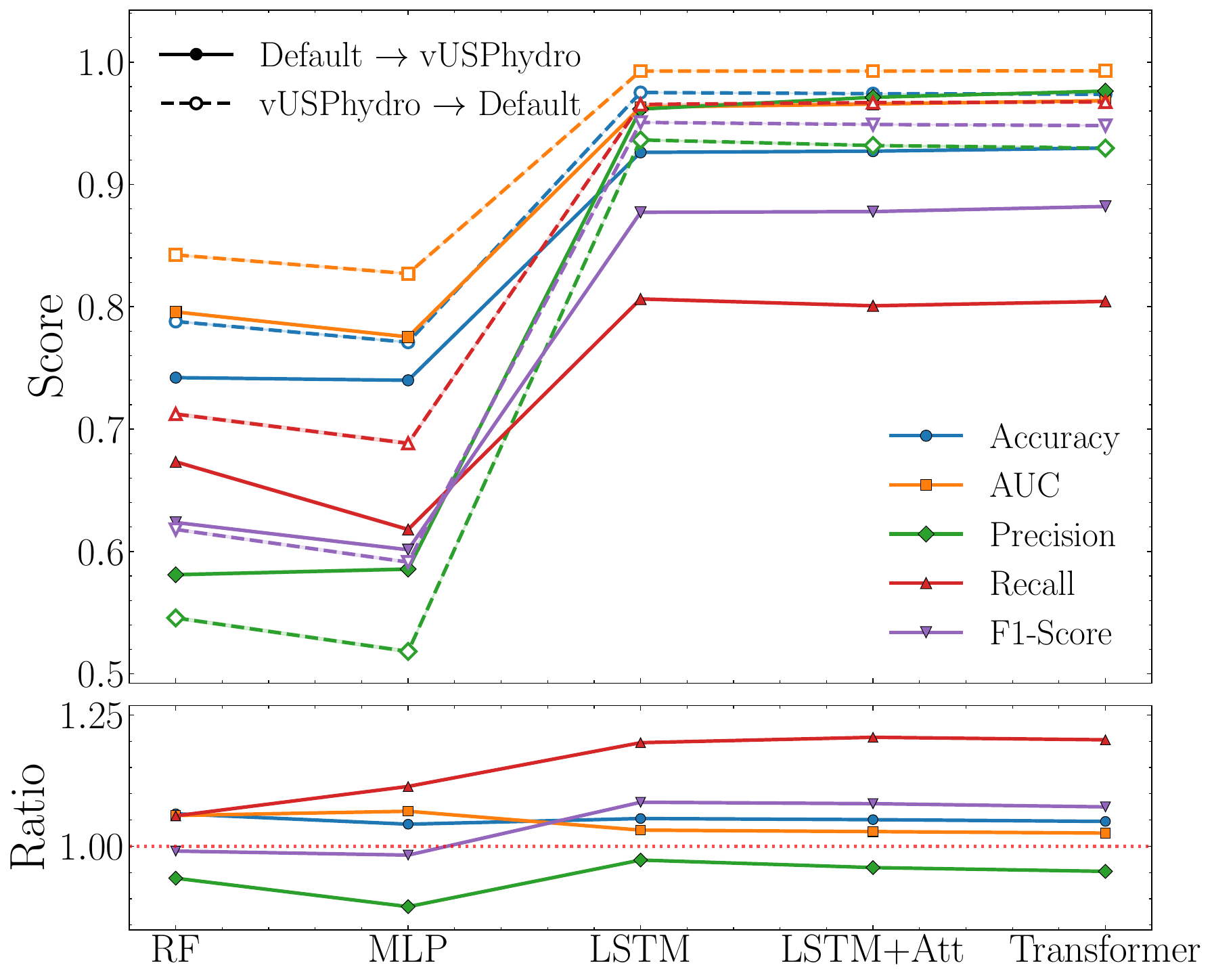}
        \caption{$p_T \in [200, 400]$ GeV}
        \label{fig:crossdomain_200_400}
    \end{subfigure}

    \caption{\justifying Cross-domain performance for each $p_T$ interval. Models are trained on one medium (Default or v-USPhydro) and tested on the other. Sequential models retain strong performance when trained on v-USPhydro, but show degradation when trained on Default and applied to the hydrodynamical medium. Here, the ratio is defined as the value of a given metric obtained with v-USPhydro $\to$ Default divided by the corresponding value obtained with the Default $\to$ v-USPhydro.}
    \label{fig:crossdomain_all}
\end{figure*}

To examine the capacity of LSTM, LSTM+Attention, and Transformer models to generalize the learned data, we conduct cross-domain evaluations, similarly to the approach used for the Random Forest and MLP models in Section~\ref{subsubsec:cross_domain}. In this setup, models are trained on data from one medium, either Default or v-USPhydro, and subsequently evaluated on data from the alternate medium. This testing directly probes the robustness and generalizability of the learned representations, offering insights into the models’ ability to adapt across simulation domains.

\begin{table}[htb]
\centering
\caption{LSTM cross-domain performance. Models trained on Default and tested on v-USPhydro, and vice-versa, for different $p_T$ intervals.}
\begin{ruledtabular}
\begin{tabular}{lccc}
$p_T$ [GeV] & Metric & Def$\rightarrow$v-USP & v-USP$\rightarrow$Def \\
\hline
\multirow{5}{*}{40--60}
    & Accuracy   & 0.9047 (0.0004) & 0.9613 (0.0003) \\
    & AUC        & 0.9621 (0.0003) & 0.9887 (0.0002) \\
    & Precision  & 0.9636 (0.0005) & 0.9066 (0.0008) \\
    & Recall     & 0.7414 (0.0010) & 0.9459 (0.0006) \\
    & F1-Score   & 0.8381 (0.0007) & 0.9258 (0.0005) \\
\hline
\multirow{5}{*}{80--250}
    & Accuracy   & 0.9212 (0.0002) & 0.9673 (0.0001) \\
    & AUC        & 0.9682 (0.0002) & 0.9916 (0.0001) \\
    & Precision  & 0.9756 (0.0002) & 0.9110 (0.0005) \\
    & Recall     & 0.7814 (0.0006) & 0.9636 (0.0003) \\
    & F1-Score   & 0.8678 (0.0004) & 0.9365 (0.0003) \\
\hline
\multirow{5}{*}{200--400}
    & Accuracy   & 0.9262 (0.0003) & 0.9751 (0.0002) \\
    & AUC        & 0.9631 (0.0003) & 0.9927 (0.0001) \\
    & Precision  & 0.9616 (0.0005) & 0.9364 (0.0006) \\
    & Recall     & 0.8064 (0.0009) & 0.9654 (0.0005) \\
    & F1-Score   & 0.8772 (0.0006) & 0.9507 (0.0004) \\
\end{tabular}
\end{ruledtabular}
\label{tab:lstm_crossdomain}
\end{table}

\begin{table}[htb]
\centering
\caption{LSTM+Attention cross-domain performance. Models trained on Default and tested on v-USPhydro, and vice-versa, for different $p_T$ intervals.}
\begin{ruledtabular}
\begin{tabular}{lccc}
$p_T$ [GeV] & Metric & Def$\rightarrow$v-USP & v-USP$\rightarrow$Def \\
\hline
\multirow{5}{*}{40--60}
    & Accuracy   & 0.8998 (0.0004) & 0.9577 (0.0003) \\
    & AUC        & 0.9602 (0.0003) & 0.9898 (0.0002) \\
    & Precision  & 0.9605 (0.0005) & 0.8824 (0.0008) \\
    & Recall     & 0.7290 (0.0010) & 0.9627 (0.0005) \\
    & F1-Score   & 0.8289 (0.0007) & 0.9208 (0.0005) \\
\hline
\multirow{5}{*}{80--250}
    & Accuracy   & 0.9217 (0.0002) & 0.9728 (0.0001) \\
    & AUC        & 0.9677 (0.0002) & 0.9919 (0.0001) \\
    & Precision  & 0.9656 (0.0003) & 0.9359 (0.0004) \\
    & Recall     & 0.7917 (0.0006) & 0.9568 (0.0003) \\
    & F1-Score   & 0.8700 (0.0004) & 0.9462 (0.0003) \\
\hline
\multirow{5}{*}{200--400}
    & Accuracy   & 0.9272 (0.0003) & 0.9742 (0.0002) \\
    & AUC        & 0.9657 (0.0003) & 0.9927 (0.0001) \\
    & Precision  & 0.9712 (0.0004) & 0.9318 (0.0007) \\
    & Recall     & 0.8008 (0.0009) & 0.9669 (0.0005) \\
    & F1-Score   & 0.8778 (0.0006) & 0.9490 (0.0004) \\
\end{tabular}
\end{ruledtabular}
\label{tab:lstmatt_crossdomain}
\end{table}

The results of these evaluations are summarized in Tables~\ref{tab:lstm_crossdomain}, \ref{tab:lstmatt_crossdomain}, and \ref{tab:transformer_crossdomain}, which present the performance of the LSTM, LSTM+Attention, and Transformer models, respectively, across different $p_T$ intervals.

Performance reveals a clear asymmetry across all sequential models. When trained on v-USPhydro and tested on Default, the models maintain high performance, showing results comparable to in-domain evaluations. In contrast, models trained on Default and evaluated on v-USPhydro suffer significant performance drops, particularly in recall and F$_1$-score. This indicates that the hydrodynamical medium enables learning of more generalizable features, while the Default scenario lacks sufficient complexity to train models capable of predicting in a more intricate scenario, as was observed in the non-sequential case.

\begin{table}[htb]
\centering
\caption{Transformer cross-domain performance. Models trained on Default and tested on v-USPhydro, and vice versa, for different $p_T$ intervals. Values are presented as mean(error).}
\begin{ruledtabular}
\begin{tabular}{llcc}
$p_T$ [GeV] & Metric     & Def$\rightarrow$v-USP & v-USP$\rightarrow$Def \\ \hline
\multirow{5}{*}{40--60}
           & Accuracy   & 0.9081 (0.0004) & 0.9617 (0.0003) \\
           & AUC        & 0.9647 (0.0003) & 0.9897 (0.0002) \\
           & Precision  & 0.9635 (0.0005) & 0.8987 (0.0008) \\
           & Recall     & 0.7524 (0.0010) & 0.9580 (0.0005) \\
           & F1-Score   & 0.8450 (0.0007) & 0.9274 (0.0005) \\ \hline
\multirow{5}{*}{80--250}
           & Accuracy   & 0.9365 (0.0002) & 0.9754 (0.0001) \\
           & AUC        & 0.9749 (0.0001) & 0.9930 (0.0001) \\
           & Precision  & 0.9791 (0.0002) & 0.9407 (0.0004) \\
           & Recall     & 0.8258 (0.0005) & 0.9622 (0.0003) \\
           & F1-Score   & 0.8959 (0.0003) & 0.9513 (0.0003) \\ \hline
\multirow{5}{*}{200--400}
           & Accuracy   & 0.9297 (0.0003) & 0.9737 (0.0002) \\
           & AUC        & 0.9685 (0.0002) & 0.9928 (0.0001) \\
           & Precision  & 0.9763 (0.0004) & 0.9297 (0.0006) \\
           & Recall     & 0.8044 (0.0009) & 0.9674 (0.0005) \\
           & F1-Score   & 0.8820 (0.0005) & 0.9481 (0.0004) \\
\end{tabular}
\end{ruledtabular}
\label{tab:transformer_crossdomain}
\end{table}

In the Transformer case, the same asymmetry persists with important nuances. When trained on v-USPhydro and evaluated on Default (v-USP$\rightarrow$Def), the model maintains excellent performance, with accuracy, AUC, and F$_1$-score remaining high, and recall consistently above 0.9 across all $p_T$ intervals. This demonstrates again that training on a realistic medium enables the Transformer with generalized representations that remain informative even in simpler environments.

Conversely, training on Default and testing on v-USPhydro (Def$\rightarrow$v-USP) leads to a performance drop, especially in recall. Despite high precision, reflecting a great classification strategy, the model fails to capture a large fraction of quenched jets. This behavior illustrates the presence of a conservative bias in the model that only labels jets as quenched when exposed to strong effects and misses valid cases with more subtle modifications.

These findings demonstrate that even advanced models such as Transformers, designed to capture long-range dependencies, cannot overcome the limitations imposed by insufficient physics in the training data. While their inductive bias allows them to attend to complex correlations across the jet evolution, such correlations must exist in the training domain to be learned effectively. This shows that realistic data is essential to ensure that ML models do not merely interpolate within idealized scenarios, but generalize meaningfully to the full complexity of jet quenching phenomena.

This result becomes especially significant when compared to the behavior of non-sequential models, the Random Forest and MLP. In those models, cross-domain performance was inferior, but it remained unclear whether this was due to their architectural limitations or genuine sensitivity to changes in the medium. The sequential models now clarify this point, showing that even with sufficient representational capacity and robust performance, the models remain sensitive to domain shifts arising from different medium descriptions. Therefore, this constitutes an important result, highlighting that machine learning models, even when powerful and well-regularized, reflect the physical assumptions embedded in the data used for training.

Figures \ref{fig:indomain_all} and \ref{fig:crossdomain_all} compare the five metrics results across all the models and medium configurations discussed in this manuscript.

In conclusion, these results confirm that sequential models, particularly the LSTM and Transformer used in this work, possess greater representational power and generalization capability compared to the chosen non-sequential architectures.

\subsubsection{Sequence Importance Analysis}

\begin{figure}[h!]
    \centering
    \includegraphics[width=1.0\linewidth]{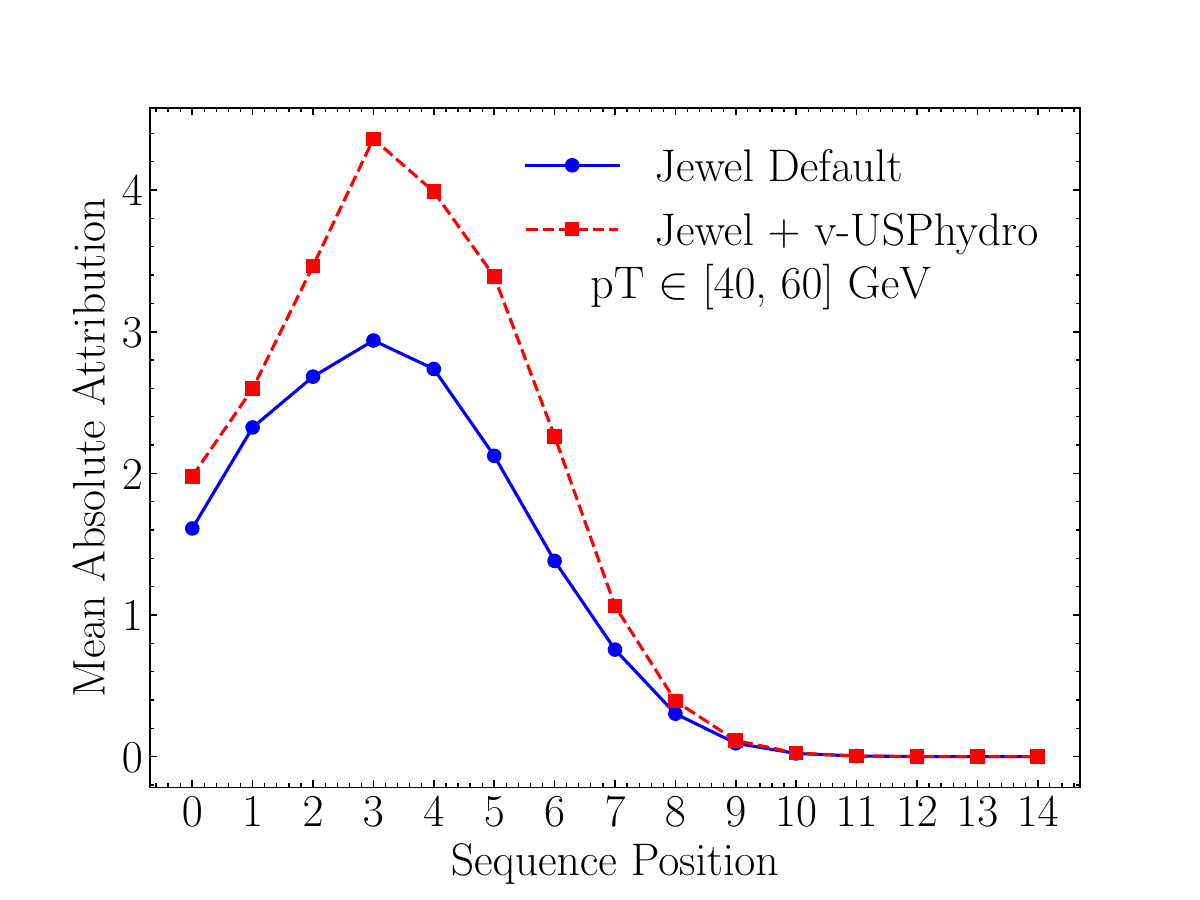}
    \caption{\justifying SHAP mean absolute attribution to each step of the grooming sequence for both \textsc{Jewel} Default and \textsc{Jewel} v-USPhydro media in the range $40 \leq p_T \leq 60$~GeV.}
    \label{fig:seq_shap_lstm_40_60}
\end{figure}

\begin{figure}[h!]
    \centering
    \includegraphics[width=1.0\linewidth]{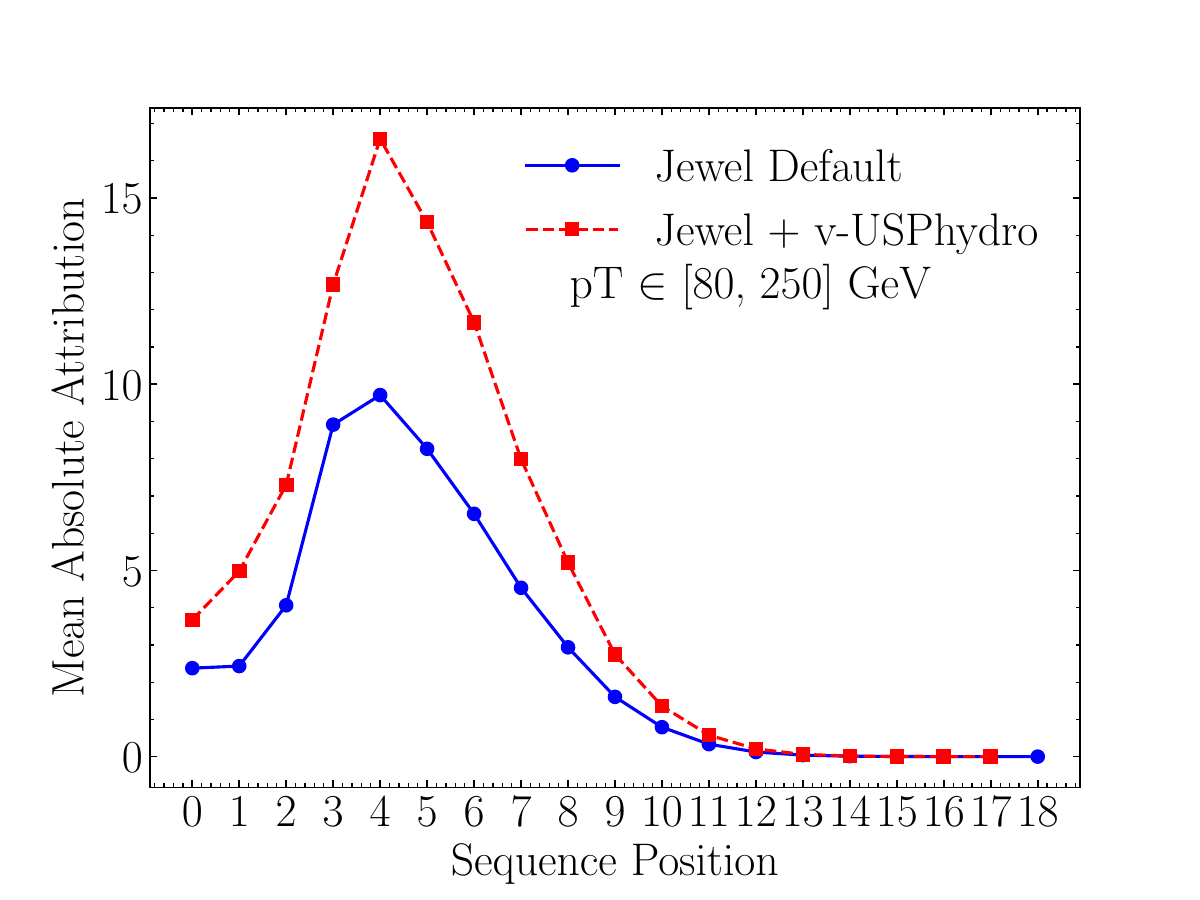}
    \caption{\justifying SHAP mean absolute attribution to each step of the grooming sequence for both \textsc{Jewel} Default and \textsc{Jewel} v-USPhydro media in the range $80 \leq p_T \leq 250$~GeV.}
    \label{fig:seq_shap_lstm_80_250}
\end{figure}

\begin{figure}[h!]
    \centering
    \includegraphics[width=1.0\linewidth]{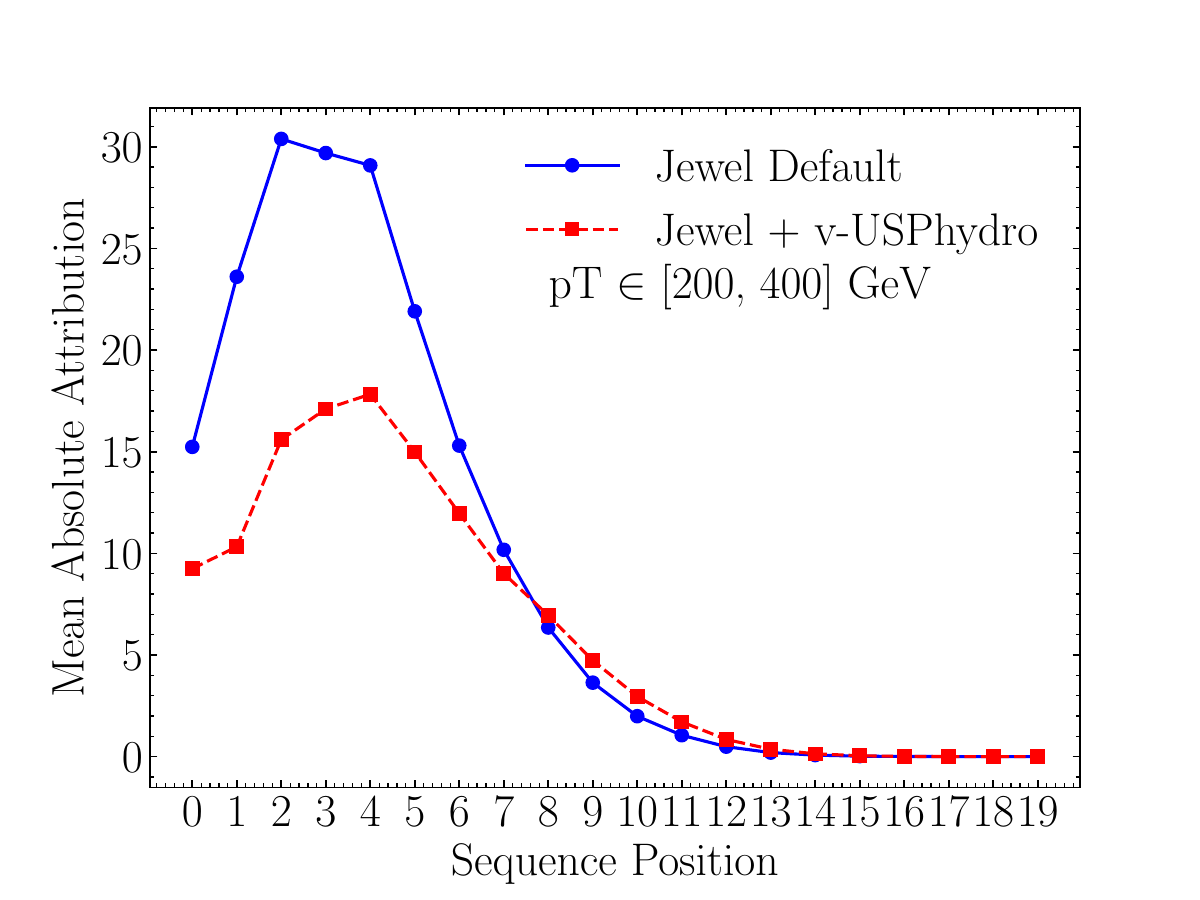}
    \caption{\justifying SHAP mean absolute attribution to each step of the grooming sequence for both \textsc{Jewel} Default and \textsc{Jewel} v-USPhydro media in the range $200 \leq p_T \leq 400$~GeV.}
    \label{fig:seq_shap_lstm_200_400}
\end{figure}

To investigate the relevance of each position in the Soft Drop sequence, we compute the absolute SHAP attribution step by step for every jet and then take the average. In this way, the curves indicate the average importance of each declustering step in the model decision. The distributions of the mean absolute SHAP value along the grooming sequence are shown in Figures~\ref{fig:seq_shap_lstm_40_60}, \ref{fig:seq_shap_lstm_80_250}, and \ref{fig:seq_shap_lstm_200_400}.


Notice that most of the attribution is concentrated in the first few Soft Drop splittings, typically around \(t = 2\text{--}5\), whereas the later steps contribute very little to the prediction. This pattern is consistently found for both media and across all \(p_T\) intervals, indicating that the network primarily relies on information encoded in the earliest and hardest branchings of the declustering sequence. %Comparing the two media within the same model, the \textsc{Jewel} v-USPhydro case exhibits a slightly broader distribution of importance over the sequence than the \textsc{Jewel} Default case, which may reflect the effect of a more complex evolution of the medium rather than a simpler one.

A more detailed comparison between the two media in the \(40 \leq p_T \leq 60\) GeV interval shows that both attribution profiles reach their maximum at \(t=4\), but with a different internal hierarchy. In the \textsc{Jewel}+v-USPhydro case, the peak at \(t=4\) is more than twice as large as the attribution of the first splitting, indicating a strong enhancement of the discriminating information after the beginning of the declustering sequence. In contrast, the \textsc{Jewel} Default profile also peaks at \(t=4\), but the increase relative to the first splitting is milder, rising only from about 2 to 3 in SHAP magnitude rather than doubling. This difference suggests that, while both media encode relevant information beyond the first declustering step, the v-USPhydro medium does so in a much more pronounced way, shifting the model sensitivity toward the next few resolved branchings. %Moreover, the v-USPhydro curve remains above the Default one throughout essentially the full relevant region \(t=1\)–7, which indicates that the evolving medium not only enhances the peak contribution, but strengthens the importance of the early sequence as a whole.

This pattern becomes even clearer in the \(80 \leq p_T \leq 250\) GeV range. In the v-USPhydro case, the peak around \(t=5\) is roughly three to four times larger than the attribution at the first splitting, while in the Default case, the enhancement from the first splitting to the maximum is also strong but noticeably less pronounced. This means that, for v-USPhydro, the model extracts most of the useful information not from the entrance of the sequence, but from a delayed region of the declustering history, after a few branchings have taken place. %Physically, this may indicate that the medium-induced signature encoded in the groomed sequence is not simply an overall softening of the jet, but a restructuring of the early branchings that becomes most visible once the shower has evolved beyond the first split. In other words, what distinguishes the v-USPhydro jets is not the presence of one exceptional splitting, but a stronger hierarchy between the first declustering and the subsequent few steps.

The high-\(p_T\) interval, \(200 \leq p_T \leq 400\) GeV, shows a qualitatively different behavior. Here, the Default medium displays a much stronger concentration of attribution in the earliest steps, with a very high plateau-like structure between \(t=3\) and \(t=5\). In contrast, the v-USPhydro curve is lower and smoother, with a more modest peak and a less dramatic separation between the leading early steps. This suggests that, at high \(p_T\), the Default medium produces a more sharply localized signature in the first few branchings, whereas v-USPhydro distributes the relevant information more evenly and with less contrast. %Therefore, the distinction between media is \(p_T\)-dependent, since at low and intermediate \(p_T\), v-USPhydro enhances the relative importance of the early (but not first splittings), while at high \(p_T\) the Default scenario leads to a more dominant and more concentrated early-sequence imprint.

These results indicate that the distinction between the Default and v-USPhydro media is \(p_T\)-dependent. At low and intermediate \(p_T\), the v-USPhydro medium leads to larger SHAP attributions over the first few declustering steps and to a slightly more extended sequence dependence. At high \(p_T\), however, the Default medium becomes more strongly concentrated in the earliest splittings, while v-USPhydro retains a broader but less sharply peaked attribution profile.

In the present work, this sequential SHAP analysis was carried out only for the LSTM architecture. Extending the same study to the LSTM with Attention and Transformer models will be an important next step, since it will allow us to test whether this dominance of the first splittings can be interpreted as a physical feature of the data or a behavior that depends on the specific bias of the model. Such a comparison may also help clarify whether architectures with explicit attention mechanisms capture the same temporal structure or redistribute the importance across the grooming sequence in a qualitatively different way.

\section*{Summary}
In this work, we employ the Monte Carlo event generator \textsc{Jewel} with two different descriptions of the hot and dense medium formed in heavy-ion collisions, such as Pb--Pb, to simulate quenching effects on jets. The main goal of this manuscript is to determine which of the chosen models of Machine Learning have greater generalization power, using observables computed only at the first splitting of the jet tree history and a sequence of observables computed within the grooming history of a jet.

We demonstrate the existence of an architectural hierarchy in machine learning approaches for jet quenching identification through systematic cross-domain analysis of medium-modified jets in heavy-ion collisions. Sequential neural networks, LSTM, LSTM+Attention and Transformer, exhibit exceptional classification performance ($\geq$ 93\% accuracy) and remarkable generalization capabilities across distinct medium implementations, while static data-based algorithms, Random Forest and Multilayer Perceptron, display dramatic performance degradation under domain transfer conditions. 

The cross-domain validation reveals a critical asymmetry, showing that models trained on the more realistic v-USPhydro hydrodynamic medium maintain robust discrimination when evaluated on the \textsc{Jewel} Default scenario, reaching accuracies of at least \(90\%\), whereas the reverse configuration leads to a pronounced degradation in performance, with accuracies dropping to around \(65\%\) for the non-sequential models. This result indicates that the more complex medium dynamics of v-USPhydro allow the models to learn a richer set of physical patterns, some of which remain identifiable even in the simpler Default setup. In this sense, the medium leaves fingerprints on the final observables, since the information learned in the realistic scenario is sufficiently structured and robust to generalize across media, while the features learned in the Default case are too limited to capture the broader phenomenology present in v-USPhydro.

It is worth noting that the differences caused by the two different media in jet observables are not immediately evident in global statistical measurements, such as \(R_{AA}\), which average over large jet collections \cite{monalisa2024}. In contrast, they become more visible in a jet-by-jet analysis based on machine learning, highlighting the potential of these methods to uncover patterns that may remain hidden in traditional ensemble observables.
%Despite this superior generalization, sequential models retain medium implementation details, indicating successful capture of subtle jet-medium interaction mechanisms rather than superficial pattern recognition. The SHAP analysis of the LSTM sequence showed emphasis on early jet stages, suggesting that medium-induced modifications manifest predominantly during initial partonic interactions. 

Interpretability analysis of non-sequential models identifies the groomed angular separation and jet mass as the primary discriminative features, with quenched jets exhibiting enhanced substructure broadening consistent with theoretical expectations. In this context, SHAP proves to be a valuable tool for gaining insight into how jets are modified while traversing a hot and dense medium. For the sequential LSTM-based analysis, the attribution patterns further show that the earliest steps of the Soft Drop sequence carry most of the relevant information for the classification task, with the importance strongly concentrated in the first few declustering stages and rapidly decreasing at later steps. Taken together, these results indicate that both the global substructure observables and the early organization of the grooming sequence encode the dominant signatures of medium modification.

%To establish a comprehensive understanding, ongoing investigations employ advanced interpretability methodologies, including SHAP decomposition, Integrated Gradients attribution, and multi-layer attention visualization, to systematically dissect Transformer decision pathways and establish quantitative correlations between learned representations and underlying Quantum Chromodynamics phenomena. 

However, our simulation framework still lacks embedded thermal background fluctuations, which represents an important limitation and may artificially enhance the classification performance. In addition, one should keep in mind that not every jet reconstructed in a Pb--Pb event is necessarily a strongly quenched jet, since some partons may traverse only a short path in the medium or undergo relatively low modification. This intrinsic physical ambiguity complicates the interpretation of any binary classification task. Future investigations incorporating realistic thermal medium fluctuations, experimental background conditions, and more comprehensive neural-network interpretability frameworks will therefore be essential to assess both the practical applicability and the physical robustness of these sequential machine-learning methodologies for jet-quenching studies in real heavy-ion collision environments.

\section*{Data Availability}

The data and analysis code supporting the findings of this study are available in a public GitHub repository \cite{Leonardo2026}.

\begin{acknowledgments}
We acknowledge the Brazilian funding agency FAPESP (Fundação de Amparo à Pesquisa do Estado de São Paulo), grant number 2024/02346-6, associated with the project 2020/04867-2, as well as the University of São Paulo (USP) for financial support.
\end{acknowledgments}

%\nocite{*}

\bibliography{apssamp}% Produces the bibliography via BibTeX.

\clearpage
\newpage

\appendix
\section{Machine Learning Hyperparameters}\label{app:hyperparameters}
%\chapter{Machine Learning Hyperparameters}\label{app:hyperparameters_app_ml}

In this appendix, we present the hyperparameters used for each model in the training and evaluation processes.  

The Random Forest model was imported from the \texttt{scikit-learn} \cite{pedregosa2011scikit} library, version 1.5.1. All the supervised neural networks used in this work were built with \textsc{PyTorch} 2.3.1 \cite{paszke2019pytorch}. Hyperparameter optimization was conducted with \texttt{Hyperopt} \cite{bergstra2013modelsearch}.

%\subsection{Supervised Models}

\subsection{Random Forest}

\begin{center}
\captionof{table}{Optimized Random Forest hyperparameters used in all $p_T$ intervals and for both medium descriptions.}
\label{tab:rf_hyperparams}
\begin{tabular}{lc}
\toprule
\textbf{Hyperparameter} & \textbf{Value} \\
\midrule
Number of estimators (\texttt{n\_estimators}) & 400 \\
Maximum depth (\texttt{max\_depth}) & 30 \\
Minimum samples per split (\texttt{min\_samples\_split}) & 10 \\
Minimum samples per leaf (\texttt{min\_samples\_leaf}) & 2 \\
Bootstrap sampling (\texttt{bootstrap}) & False \\
Maximum features (\texttt{max\_features}) & \texttt{sqrt} \\
Random seed (\texttt{random\_state}) & 42 \\
Number of jobs (\texttt{n\_jobs}) & 1 \\
\bottomrule
\end{tabular}
\end{center}

\subsection{MLP}

\begin{center}
\captionof{table}{Optimized MLP hyperparameters for the $p_T \in [40,60]$~GeV interval.}
\label{tab:mlp_40_60}
\begin{tabular}{lcc}
\toprule
\textbf{Hyperparameter} & \textbf{Default} & \textbf{v-USPhydro} \\
\midrule
Decay factor   & 0.974 & 0.945 \\
Dropout rate   & 0.329 & 0.129 \\
Hidden size 1  & 64    & 16 \\
Hidden size 2  & 12    & 16 \\
Learning rate  & 0.003 & 0.003 \\
Loss function  & BCE   & BCE \\
Batch size     & 64    & 128 \\
Epochs         & 40    & 60 \\
\bottomrule
\end{tabular}
\end{center}

\begin{center}
\captionof{table}{Optimized MLP hyperparameters for $p_T \in [80,250]$ and $[200,400]$~GeV intervals.}
\label{tab:mlp_highpt}
\begin{tabular}{lcc}
\toprule
\textbf{Hyperparameter} & \textbf{Default} & \textbf{v-USPhydro} \\
\midrule
Decay factor   & 0.945 & 0.936 \\
Dropout rate   & 0.129 & 0.235 \\
Hidden size 1  & 16    & 32 \\
Hidden size 2  & 16    & 12 \\
Learning rate  & 0.003 & 0.001 \\
Loss function  & BCE   & BCE \\
Batch size     & 128   & 64 \\
Epochs         & 60    & 70 \\
\bottomrule
\end{tabular}
\end{center}

\subsection{LSTM and LSTM+Attention}
\begin{center}
\captionof{table}{Optimized LSTM and LSTM+Attention hyperparameters used for all $p_T$ intervals and for both medium descriptions.}
\label{tab:lstm_hyperparams}
\begin{tabular}{lc}
\toprule
\textbf{Hyperparameter} & \textbf{Value} \\
\midrule
Decay factor          & 0.947 \\
Dropout rate          & 0.456 \\
Hidden size 1         & 18 \\
Hidden size 2         & 6 \\
Learning rate         & 0.016 \\
Loss function         & BCE \\
Batch size            & 64 \\
Epochs                & 40 \\
Number of LSTM layers & 2 \\
\bottomrule
\end{tabular}
\end{center}

\subsection{Transformer}

\begin{center}
\captionof{table}{Optimized Transformer hyperparameters used for all $p_T$ intervals and for both medium descriptions.}
\label{tab:transformer_hyperparams}
\begin{tabular}{lc}
\toprule
\textbf{Hyperparameter} & \textbf{Value} \\
\midrule
Batch size                   & 256 \\
Decay factor                 & 0.936 \\
Dropout rate                 & 0.101 \\
Feed-forward dimension       & 256 \\
Learning rate                & 0.0013 \\
Model (embedding) dimension  & 64 \\
Epochs                       & 40 \\
Number of attention heads    & 4 \\
Number of Transformer layers & 5 \\
\bottomrule
\end{tabular}
\end{center}

\end{document}